\documentclass[iop]{emulateapj}
\usepackage{graphicx,amsmath,amsthm,amssymb,multirow,color}

\newcommand{\cmsq}{\hbox{cm$^{-2}$}}

\newcommand{\lumin}{\hbox{erg~s$^{-1}$}}
\newcommand{\lumind}{\hbox{erg~s$^{-1}$~Hz$^{-1}$}}

\newcommand{\aox}{$\alpha_{\rm ox}$}
\newcommand{\nh}{\hbox{${N}_{\rm H}$}}
\newcommand{\msun}{\hbox{${M}_{\odot}$}}

\newcommand{\msig}{M--$\sigma$}

\newcommand{\be}{\begin{equation}}
\newcommand{\ee}{\end{equation}}
\newcommand{\ba}{\begin{eqnarray}}
\newcommand{\ea}{\end{eqnarray}}

\newcommand{\chandra}{\emph{Chandra}}

\newcommand{\simgt}{\lower 2pt \hbox{$\, \buildrel {\scriptstyle >}\over {\scriptstyle\sim}\,$}}
\newcommand{\simlt}{\lower 2pt \hbox{$\, \buildrel {\scriptstyle <}\over {\scriptstyle\sim}\,$}}
\newcommand{\ls}{\lower 2pt \hbox{$\;\scriptscriptstyle \buildrel<\over\sim\;$}}
\newcommand{\gs}{\lower 2pt \hbox{$\;\scriptscriptstyle \buildrel>\over\sim\;$}}

\definecolor{Mygrey}{gray}{0.75}

\begin{document}

\def\arcsec{$^{\prime\prime}$}
\def\arcmin{$^{\prime}$}
\def\degr{$^{\circ}$}

\title[Unveiling the X-ray Properties of BALQSOs]{Unveiling the Intrinsic X-ray Properties of Broad Absorption Line Quasars with a Relatively Unbiased Sample}
\author{Leah K. Morabito\altaffilmark{1}$^,$\altaffilmark{2}, Xinyu Dai\altaffilmark{1}, Karen M. Leighly\altaffilmark{1}, Gregory R. Sivakoff\altaffilmark{3}, Francesco Shankar\altaffilmark{4}}
\altaffiltext{1}{Homer L. Dodge Department of Physics \& Astronomy, University of Oklahoma, Norman, OK, USA, }
\altaffiltext{2}{Leiden Observatory, P.O. Box 9513, 2300 RA, Leiden, The Netherlands}
\altaffiltext{3}{Department of Physics, University of Alberta, CCIS 4-183 Edmonton, AB T6G 2E1, Canada}
\altaffiltext{4}{School of Physics and Astronomy, University of Southampton, Southampton SO17 IBJ, England}

\begin{abstract}
There is growing evidence of a higher intrinsic fraction of broad absorption line quasars (BALQSOs)
than that obtained in optical surveys, on which most previous X-ray studies of BALQSOs have focused. Here we present Chandra observations of 18 BALQSOs at $z\sim2$, selected from a near-infrared (2MASS) sample, where the BALQSO fraction is likely to be close to the intrinsic fraction. We measure photon indices using the stacked spectra of the optically-faint ($i-K_s\geq 2.3$ mag) and optically-bright ($i-K_s < 2.3$ mag) samples to be $\Gamma \simeq 1.5$--$2.1$. We constrain their intrinsic column density by modelling the X-ray fractional hardness ratio, finding a mean column density of $3.5\times10^{22}$ \cmsq\ assuming neutral absorption. We incorporate SDSS optical measurements (rest frame UV) to study the broadband spectral index between the X-ray and UV bands, and compare this to a large sample of normal quasars. We estimate that the optically-faint BALQSOs are X-ray weaker than the optically-bright ones, and the entire sample of BALQSOs are intrinsically X-ray weak when compared to normal AGN. Correcting for magnification of X-ray emission via gravitational lensing by the central black hole viewed at large inclination angles makes these BALQSOs even more intrinsically X-ray weak.
Finally, we estimate AGN kinetic feedback efficiencies of a few percent for an X-ray wind of $0.3c$ in high-ionization BALQSOs. Combined with energy carried by low-ionization BALQSOs and UV winds, the total kinetic energy in BALQSOs can be sufficient to provide AGN kinetic feedback required to explain the
co-evolution between black holes and host galaxies.
\end{abstract}


\section{Introduction}
Broad absorption line quasars are a sub-sample of quasars exhibiting blue-shifted absorption troughs. In geometric models of quasars, BALQSOs are quasars viewed at large inclination angles close to the equatorial plane \citep[e.g.,][]{wey91,ogle99,sch99} or the polar direction \citep{zhou06,ghosh07}. In these models, the fraction of BALQSOs constrains the opening angle of the BAL wind. Evolutionary models place BALQSOs at the early stages of quasar evolution \citep[e.g.,][]{haz84,sh87,bm92,becker00}, where the BAL fraction measures the fraction of time when quasars are in the BAL stage. Recent studies suggest hybrid BALQSO models that combine different geometric and evolutionary ingredients \citep{bh10, dai12, di13} are needed to better explain the differences and similarities between subsamples of BALQSOs. Understanding the properties of BALQSOs is therefore important to understanding the properties of quasars in general. 

Studies of BALQSOs have typically focused on optically-selected samples. However, BALQSOs exhibit strong absorption and more dust extinction than normal quasars \citep{spray92,reichard03}, which provide sources of obscuration that make them more difficult to detect in optical surveys. The NIR provides a window where there is little spectral difference between BALQSOs and non-BALQSOs. Measuring the number of BALQSOs in optical samples provides a raw fraction in the range of $10-26$\% depending on the traditional or relaxed definition \citep[e.g.,][]{wey91,tolea02,reichard03}. However, using the near-IR instead of optical, \citet{dai08} found that the BAL fraction doubles, and ranges from $10\pm2-43\pm2$\%, confirming the corrected fraction of \citet{hf03}. \citet{dai08} modelled the fraction of BALQSOs from the SDSS-g to 2MASS-Ks band using the obscuration model, and the results were consistent with the observations, indicating that the optical BAL fraction is significantly biased. Subsequently, those results have been confirmed by several studies including \citet[][using the same data set but a different BAL definition]{ganguly08}, \citet[][using a different IR data set but the same BAL definition]{maddox08}, \citet[][using radio data]{s08}, and direct modelling of the correction factor using the optical data only \citep{knigge08}. Therefore, the 2MASS sample of BALQSOs should provide a relatively unbiased sample for studying various properties of BALQSOs. 

A BALQSO is traditionally defined by a BAL wind trough at least 2000 km s$^{-1}$ wide \citep{wey91}, although a more relaxed definition \citep[1000 km s$^{-1}$ trough at least,][]{trump06}, is also used. The BALQSOs  are divided into two classes: those with absorption troughs from low ionization species are low ionization BALQSOs (LoBALs), and those with absorption troughs from high ionization species are high ionization BALQSOs (HiBALs). All LoBALs also have high ionization troughs in their spectra. The majority of BALQSOs are classified only as HiBALs, and our NIR-selected sample is representative of this population. 

Another distinct property of these objects is their apparent X-ray weakness compared to normal quasars, which is attributed to intrinsic absorption with $N_H$ column densities of $10^{22-24}$ \cmsq\ \citep[e.g.,][]{gall99,gall02,gall06,green01,chartas01,grupe03,stalin11}.  These studies generally conclude that, after correcting the intrinsic absorption in BALQSOs, the broad band UV to X-ray spectral energy distributions are consistent between BALQSOs and normal quasars. 
Several other studies, however, find that BALQSOs are X-ray weak \citep[e.g.,][]{sh01,clavel06,grupe08}, although a large fraction of these objects are low-ionization BALQSOs. An important effect, which was neglected in almost all previous studies of BALQSOs, is the gravitational lensing effect of X-ray emission by the central black hole.  Recently, microlensing and X-ray reverberation mapping methods have constrained X-ray emission of AGN to $\sim 10 r_g$ \citep{dai10, bl11, bl13, morgan12, mosquera13, rm13} including a measurement for a mini-BALQSO \citep{morgan08}, where $r_g = GM/c^2$ is the gravitational radius of an object.
For X-ray emission of this size, the gravitational lensing effect of the central black hole will produce strong inclination-dependent flux profiles even if the unlensed emission is isotropic \citep[e.g.,][]{chen13a, chen13b}, where the X-ray emission viewed from large inclination angles is strongly magnified.
This effect is important for comparing objects viewed from different inclination angles such as BALQSOs and non-BAL quasars, assuming the unified model of AGN \citep[e.g.,][]{ant93}. 

Under the paradigm of the co-evolution of AGN and the host galaxies, AGN feedback is widely used in almost all models of galaxy formation \citep[e.g.,][]{granato04,so04,hopkins05}, that reproduce several observables such as the galaxy stellar mass function and the \msig\ relation. A minimum feedback efficiency, $\epsilon=\dot{E_k}/L_{\text{Bol}}$, of 0.5--5\% is needed in all these models \citep[e.g.,][and references therein]{he10}. The powerful radio jets in AGNs can provide enough kinetic energy \citep[e.g.,][]{s08b}; however the fraction of luminous, radio-loud AGNs is only 10\% \citep{jiang07} and the radio-loud fraction only increases in low-luminosity AGNs \citep[e.g.,][]{lafranca10}.
In addition, radio-loud AGNs may not be an evolutionary sequence of quasars, but a different population altogether \citep[e.g.,][]{shankar10} and the coupling between the jet energy and ISM in the host galaxy may not be efficient before the jet is decelerated. Thus this process is arguably not the major galaxy-scale feedback mechanism for all AGNs.

The AGN wind is another candidate for the feedback. Recent studies show that the feedback energy from the AGN wind in Seyferts ranges from four to five orders of magnitude smaller than \citep[e.g.,][]{mathur09} to consistent with the theoretical value for the feedback \citep[e.g.,][]{ck12}. Quasar winds in BALQSOs are much stronger, and recent studies show that the feedback efficiency can reach 1--5\% from the UV absorbers in BALQSOs \citep[e.g.,][]{moe09,bo13,arav13}, consistent with the lower limit of the theoretical requirement to explain galaxy formation, although there may be a selection bias characterizing these objects \citep{lucy13}. The kinetic feedback energy from the AGN wind can be calculated as $\dot{E_k}=\dot{M}v^2/2=2\pi\mu m_p f_c r N_H v^3$ \citep{crenshaw03,moe09}, where $f_c$ (covering fraction), $r$ (location), $N_H$ (column density), and $v$ (velocity) are all properties that must be measured. With the detection of near relativistic outflows ($0.3-0.7c$) in X-ray absorption lines \citep[e.g.,][]{chartas02}, the X-ray absorbers in BALQSOs can surpass the UV absorbers, and provide sufficient feedback energy with $\epsilon\sim18-170$\% \citep{chartas09}, assuming that these absorbers are present in all BALQSOs. 

In this paper, we measure average X-ray properties including the intrinsic absorption column density in BALQSOs in an unbiased way, using the sample of the Two Micron All Sky Survey \citep[2MASS;][]{2mass} selected BALQSOs, where the intrinsic fraction of BALQSOs inferred from this sample is twice the raw fraction measured from the optical surveys \citep{dai08}. We further estimate the AGN feedback efficiency in BALQSOs as an indicator whether or not they would be good candidates for further AGN feedback studies. 

In \S~\ref{sec:dr}, we outline the sample, observations, and data reductions. This is followed by spectral analysis in \S~\ref{sec:spec}. X-ray hardness ratio calculations are described in \S~\ref{sec:hr}, and an analysis of the X-ray properties are presented in \S~\ref{sec:xray}. A comparison of the samples is presented in \S~\ref{sec:civ}, and lastly we present a discussion and conclusions in \S~\ref{sec:dis} and \S~ \ref{sec:conc}, respectively.

Throughout this paper, we use a standard cosmology of $H_0=70$ km s$^{-1}$ Mpc$^{-1}$, $\Omega_M=0.3$, $\Omega_{\Lambda}=0.7$, and $k=0$. For conciseness, after the BALQSOs are introduced in Table~\ref{tab:obs}, we will refer to individual objects by abbreviated SDSS designators, e.g., J004613.54$+$010425.7 will be referred to as J0046$+$0104.

\section{Sample Selection, Observations, and Data Reduction}\label{sec:dr}
\begin{figure}
\begin{center}
\plotone{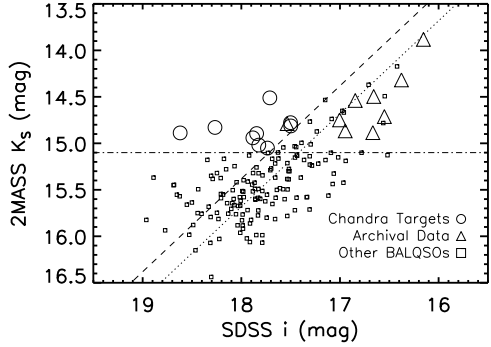}
\caption[Target Selection Plot]{Target selection plot from the HiBAL sample of \citet{gib09a} satisfying the traditional definition of \citet{wey91} within the redshift range of $1.7 < z < 2.15$. We selected BALQSOs brighter than the 99\% database completeness limit of the 2MASS survey, $K_s < 15.1$ (the horizontal dot-dash line). The dotted line ($i-K_s = 2.3$~mag) distinguishes between optically-faint (left) and optically-bright (right) BALQSOs. Eight of the optically-bright BALQSOs and one of the optically-faint BALQSO have archival \chandra\ data, and we observed nine new targets left of the dashed line ($i-K_s \ge 2.6$~mag) with \chandra. \label{fig:tgts}}
\end{center}
\end{figure}

We selected targets from the SDSS DR5 BALQSO catalog \citep{gib09a} with 2MASS $K_s < 15.1$ mag (the 99\% database completeness limit of  2MASS) in the redshift range of $1.7<z<2.15$ (Figure~\ref{fig:tgts}).  This selection yields BALQSO fractions consistent with the underlying intrinsic fractions \citep{dai08,dai12}.  The \citet{gib09a} sample used the traditional BALQSO definition \citep{wey91} of a BAL wind trough at least 2000 km s$^{-1}$ wide.  The redshift range of $1.7<z<2.15$ ensures that both the \ion{C}{4} and \ion{Mg}{2} troughs are in the SDSS spectral range, which allows us to pick objects classified by \citet{gib09a} as HiBALs. We examined the SDSS spectra and concluded that all objects are HiBALs except J2252$-$0841, which is a LoBAL and possibly an FeLoBAL. 
We further separate the whole sample into ``optically-faint'', $i-K_s\geq2.3$ mag, and ``optically-bright'', $i-K_s<2.3$ mag, samples, as shown in Figure~\ref{fig:tgts}. 
Prior to 2011, nine objects had archival \chandra\ data available.

To complete a full sample of 2MASS selected BALQSOs, we observed nine of the optically-faint BALQSOs with ACIS-S \citep{garmire03} on board \chandra\ \citep{wei02} in 2011.
General properties of the targets and \chandra\ observations are given in Table~\ref{tab:obs}.

\begin{deluxetable*}{lccrcrccl}
\tablewidth{0pt}
\tablecaption{Target List}
\tablehead{\colhead{Object} & \colhead{z} & \colhead{Gal. $N_H$\tablenotemark{a}} & \chandra & \colhead{Date-Obs} & \colhead{Exp. time} & \colhead{Net} & \colhead{Class.} & \colhead{Sample} \\
\colhead{} & \colhead{} & \colhead{[$10^{20}$~cm$^{-2}$]} & obsid & \colhead{} & \colhead{[seconds]} & \colhead{Photons} & \colhead{} & \colhead{}}
\startdata
J004613.54$+$010425.7   & 2.15 & 2.46 & 12764 & 2011-12-01 & 11096  & $8.74_{-2.71}^{+3.35}$ & Hi  & Opt. Faint \\[2pt]
J092913.96$+$375742.9   & 1.92 & 1.40 & 9162  & 2007-12-28 &  3987 &  $43.95_{-6.27}^{+6.99}$ &  Hi                      & Opt. Faint \\[2pt]
J093514.71$+$033545.7   & 1.82 & 3.64 & 5705  & 2005-03-07 &  1782  & $4.98_{-1.94}^{+2.62}$ &  Hi                      & Opt. Bright   \\[2pt]
J095929.88$+$633359.8   & 1.85 & 2.67 & 5702  & 2004-12-28 &  1526  &  $9.97_{-2.86}^{+3.53}$ &  Hi                      & Opt. Bright   \\[2pt]
J100711.81$+$053208.9   & 2.14 & 2.37 & 6882  & 2006-06-25 &  1030  & $14.99_{-3.57}^{+4.22}$  &  Hi                      & Opt. Bright  \\[2pt]
J101012.65$+$560520.5\tablenotemark{b}   & 2.14 & 7.95 & 12765 & 2011-09-07 & 9313   & $3.69_{-1.73}^{+2.38}$ & Hi                       & Opt. Faint \\[2pt]
J104130.17$+$000118.8   & 2.07 & 4.26 & 12766 & 2011-02-01 & 10304  & $57.56_{-7.54}^{+7.62}$   & Hi                       & Opt. Faint \\[2pt]
J110505.15$+$111541.0   & 1.71 & 2.25 & 12767 & 2011-04-16 & 6950   &  $3.67_{-1.72}^{+2.38}$ & Hi                       & Opt. Faint \\[2pt]
J111316.42$+$091439.0\tablenotemark{b}   & 1.71 & 2.78 & 6884  & 2006-10-28 & 1602   &  $0.99_{-0.73}^{+1.49}$ & Hi    & Opt. Bright \\[2pt]
J115944.82$+$011206.9   & 2.00 & 2.06 & 9158  & 2007-12-28 &  3706  & $153.93_{-12.29}^{+12.41}$                 &  Hi                      & Opt. Bright  \\[2pt]
J120550.19$+$020131.5   & 2.13 & 1.89 & 5700  & 2006-02-22 &  2085  &  $10.98_{-2.99}^{+3.68}$ &  Hi                      & Opt. Bright   \\[2pt]
J131028.13$+$482204.8   & 1.74 & 1.07 & 12768 & 2011-08-30 & 9910   & $22.74_{-4.47}^{+5.13}$  & Hi   & Opt. Faint  \\[2pt]
J143752.75$+$042854.5\tablenotemark{b}   & 1.92 & 2.58 & 5704  & 2005-03-22 & 1533   &  $1.00_{-0.73}^{+1.49}$ & Hi      & Opt. Bright \\[2pt]
J144707.41$+$520340.0   & 2.06 & 1.75 & 12769 & 2011-12-22 & 9221   & $47.46_{-6.65}^{+7.21}$ & Hi                       & Opt. Faint \\[2pt]
J160202.39$+$401301.3   & 2.12 & 1.29 & 12770 & 2011-11-14 & 8921   &  $5.69_{-2.16}^{+2.81}$               & Hi                       &  Opt. Faint  \\[2pt]
J162143.78$+$355533.9   & 2.05 & 1.28 & 12771 & 2011-11-28 & 11391  & $18.41_{-4.06}^{+4.70}$      & Hi                       & Opt. Faint  \\[2pt]
J225257.62$-$084141.2   & 2.15 & 3.53 & 12772 & 2011-05-16 & 9814   &  $11.49_{-3.17}^{+3.80}$ & Lo, Fe? & Opt. Faint  \\[2pt]
J231324.46$+$003444.5\tablenotemark{b}   & 2.08 & 4.08 & 5701  & 2006-03-26 & 1149   &  $1.98_{-1.13}^{+1.85}$ &  Hi     & Opt. Bright 
\enddata
\tablecomments{Classification scheme: Hi=HiBAL, Lo=LoBAL, Fe=FeLoBAL.}
\tablenotetext{a}{Gal. $N_H$ is the weighted value of the Galactic absorption column density from \citet{dl90}.}  \tablenotetext{b}{Non-detection.}
\label{tab:obs}
\end{deluxetable*}

All data were reprocessed with CIAO, using CALDB version 4.5.3, and the images were refined with \verb+subpix.pro+ \citep{mori01} to improve the spatial resolution at a sub-pixel level. Of the ten optically-faint BALQSOs shown in Figure~\ref{fig:tgts}, only one target, J1010$+$5605, was not detected via \verb+wavdetect+ when setting the significance threshold at $10^{-6}$ to ensure less than one false source per 1024x1024 field. Three of the optically-bright targets were also not detected using the same significance threshold. Background annuli with 10\arcsec\ (inner) and 20\arcsec\ (outer) radii were chosen and visually inspected in \verb+ds9+ for possible contamination from serendipitous sources. The CIAO routine \verb+mkpsfmap+ was used to find the radius enclosing 95\% of counts for a point source at each source location for the effective energies of the standard \chandra\ soft, medium, hard, and broad bands (0.92, 1.56, 2.3, and 3.8 keV, respectively) for use as source extraction regions. For all detected targets, we used \verb+aprates+ in conjunction with un-normalized exposure maps to find the photon flux in each of the \chandra\ bands. 

\section{Spectral Analysis}\label{sec:spec}
We extracted the spectra using the corresponding redistribution matrix and ancillary response files of all detected targets with the CIAO tool \verb+specextract+. Since the spectra of individual sources typically have low signal-to-noise ratios (S/N), we stacked spectra with \verb+combine_spectra+ to better constrain the spectral properties.
In particular, we stacked the spectra of the detected BALQSOs in three groups: the total, optically-faint and optically-bright samples. 
We examined whether the stacks are dominated by individual bright sources (see Table~\ref{tab:obs} for net photons), and excluded the dominant spectrum of J1159$+$0112 for separate analysis. There are five optically-bright and nine optically-faint BALQSOs in the final stacks. The spectra are displayed in Figure 2.

We fit a suite of models to four spectra: the combined spectra for the total, optically-bright, and optically-faint samples, and the individual spectrum of J1159$+$0112, using \verb+XSPEC+. All models included three components: Galactic absorption, intrinsic (redshifted) absorption, and an intrinsic emission
component. For consistency with past studies, Galactic absorption (and intrinsic absorption in some of the models) is modelled with photo-electric absorption using the \citet{mm83} cross-sections (\verb+wabs+ and \verb+zwabs+ ).We used the Cash statistic in the spectral fits. For the combined spectra, we used the appropriate geometric means of the Galactic absorption column densities from \citet{dl90} and the redshifts in the spectral fits. We fit the spectra between the observed 0.4--8 keV band, corresponding to the rest-frame of 1.2--24 keV band considering our mean redshift of $z=2$.  We list the results in Table~\ref{tab:spec}.

We first fit the spectra with a power-law modified by intrinsic and Galactic absorptions (Model I: \verb+wabs*zwabs*zpower+).
The photon indices obtained from this model ranges from $\Gamma=1.5$--2.1 for the three combined spectra, broadly consistent with mean values, $\Gamma=1.7$--1.9, for normal quasars \citep[e.g.,][]{rt00, dai04, saez08}.  We constrained the intrinsic absorption from $\nh=1.1$--1.4$\times10^{22}\cmsq$ for the three combined spectra.
We also fit the combined spectra with more complex partial covering (Model II: \verb+wabs*zpcfabs*zpower+), warm absorber (Done et al. 1992, Model III: \verb+wabs*absori*zpower+), and reflection models (Magdziarz and Zdziarski 1995, Model IV: \verb+wabs*zwabs*pexrav+), and generally obtained fitting statistics comparable to Model I.
Because of the relative low S/N, we had to fix several model parameters when fitting the warm absorber or reflection models.  These parameters include the temperature of the absorber and Fe abundance in the warm absorber model, and the abundances, power law cut-off energy, photon index and inclination angle in the reflection model (see Table~\ref{tab:spec}). 
Therefore, we cannot distinguish between these models in our spectral analysis, and we adopt the simple power law model modified by Galactic and intrinsic absorption (Model I) as our reference model for subsequent analysis.

For J1159$+$0112, we obtained a very flat photon index, $\Gamma=1.1\pm0.2$, using Model I.  This BALQSO has a 2--10 keV luminosity of $5\times10^{45}$\lumin, which predicts $\Gamma \simeq 2$, based on the $\Gamma$--$L_X$ relation \citep{dai04,saez08}.  The partial covering and warm absorber models also yield flat photon indices, and the reflection model is the only model that yields a photon index $\Gamma=1.6\pm0.5$ consistent with $\Gamma=2$.
We conclude that a reflection component is likely present in the X-ray spectrum of J1159$+$0112. 
We further fixed $\Gamma=1.9$ in the reflection model to better constrain other parameters in the model, and we constrained the reflection strength $R = 11_{-5}^{+9}$, indicating a 2$\sigma$ detection of the reflection component.

Comparing the optically-bright and optically-faint samples, we find the best-fit photon indices for the optically-faint sample are generally flatter in the various models than for the optically-bright sample.
Although the S/N for the stacked spectrum of the optically-bright sample is low in our analysis, comparison with previous optically selected samples is valid. \citet{gall02} found a mean $\Gamma \simeq 2$ for optically-selected BALQSOs, which is consistent with our result for the optically-bright sample. The steeper spectral index for the optically-bright sample may suggest more prominent complex spectral features, such as the reflection component, than in the optically-faint sample.  However, this difference is only marginally significant, 2$\sigma$, because of the large uncertainties. 

\begin{figure}
\begin{center}
\plotone{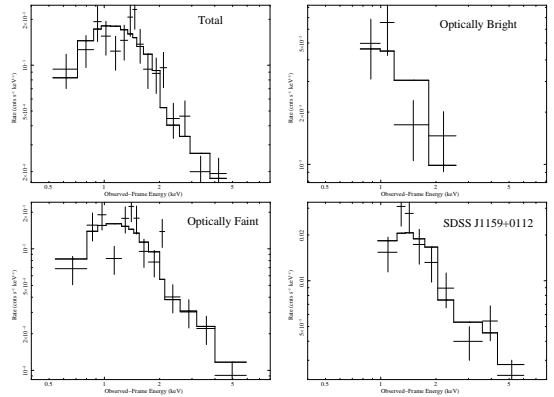}
\caption{\chandra\ spectra and best-fit models of 2MASS selected BALQSOs.  The spectra are binned for clarity, while the fitting is performed to the un-binned spectra.  The combined total, optically-bright, and optically-faint spectra are fit with a simple power law model modified by intrinsic and Galactic absorption, and the spectrum of J1159$+$0112 is fit by including an additional reflection component. \label{fig:spec}}
\end{center}
\end{figure}

\begin{deluxetable*}{lcccccc}
\tablewidth{0pt}
\tablecaption{ Spectral Fitting Results \label{tab:spec}}
\tablehead{\colhead{Sample} & \colhead{\hbox{${N}_{\rm H, intrinsic}$}} & \colhead{$\Gamma$} & \colhead{Covering Fraction} & \colhead{Xi} & \colhead{rel\_refl} & \colhead{C-Stat(dof)}\\
 &\colhead{[$10^{22}~\cmsq$]} & & & & & 
}
\startdata
\\ \tableline\tableline\\[-0.5em]
\multicolumn{7}{c}{Model I: Powerlaw with Neutral Absorption}\\
\multicolumn{7}{c}{\tt wabs*zwabs*zpower}\\[0.25em]
\tableline\tableline\\
Total            & $1.1\pm0.6$ & $1.60\pm0.15$ & \nodata & \nodata & \nodata & 375.4(516)\\
Optically Faint  & $1.3\pm0.8$ & $1.54\pm0.18$ & \nodata & \nodata & \nodata & 359.7(516)\\
Optically Bright & $1.4\pm1.3$ & $2.1\pm0.4$ & \nodata & \nodata & \nodata & 145.4(516)\\
J1159$+$0112     & $4.5\pm1.8$ & $1.1\pm0.2$ & \nodata & \nodata & \nodata & 337.0(516)\\
\\ \tableline\tableline\\[-0.5em]
\multicolumn{7}{c}{Model II: Powerlaw with Partial Covering Absorption}\\
\multicolumn{7}{c}{\tt wabs*zpcfabs*zpower}\\[0.25em]
\tableline\tableline\\
Total            & $5.5_{-4.6}^{+8.8}$ & $1.8\pm0.3$ & $0.64_{-0.24}^{+0.36}$ & \nodata & \nodata & 374.5(515) \\[2.5pt]
Optically Faint  & $7.9_{-5.0}^{+7.9}$ & $1.8\pm0.3$ & $0.71_{-0.19}^{+0.11}$ & \nodata & \nodata & 357.6(515) \\[2.5pt]
Optically Bright & $1.4_{-0.7}^{+4.0}$ & $2.1\pm0.4$ & $0.95_{-0.75}^{+0.05}$ & \nodata & \nodata & 145.5(515) \\[2.5pt]
J1159$+$0112     & $4.4_{-1.4}^{+1.9}$ & $1.1\pm0.2$ & $0.99_{-0.15}^{+0.01}$ & \nodata & \nodata & 336.9(515) \\
\\ \tableline\tableline\\[-0.5em]
\multicolumn{7}{c}{Model III: Powerlaw with Warm Absorber}\\
\multicolumn{7}{c}{{\tt wabs*absori*zpower}\tablenotemark{a}}\\[0.25em]
\tableline\tableline\\
Total            & $\phn8.6_{-\phn7.3}^{+\phn8.3}$          & $1.7\pm0.2$ & \nodata & $560_{-\phn550}^{+4380}$                & \nodata & 374.2(515) \\[2.5pt]
Optically Faint  & $12.7_{-\phn8.2}^{+10.3}$                & $1.8\pm0.2$ & \nodata & $710_{-\phn530}^{+\phn840}$             & \nodata & 357.5(515) \\[2.5pt]
Optically Bright & $\phn1.4_{-\phn0.7}^{+\phn7.9}$          & $2.1\pm0.5$ & \nodata & $\phn\phn0_{-\phn\phn\phn0}^{+\phn650}$ & \nodata & 145.4(515) \\[2.5pt]
J1159$+$0112     & $15\phd\phn_{-12\phd\phn}^{+16\phd\phn}$ & $1.2\pm0.2$ & \nodata & $\phn\phn0_{-\phn\phn\phn0}^{+1300}$    & \nodata & 336.2(515) \\
\\ \tableline\tableline\\[-0.5em]
\multicolumn{7}{c}{Model IV: Reflection with Neutral Absorption}\\
\multicolumn{7}{c}{{\tt wabs*zwabs*pexrav}\tablenotemark{b}}\\[0.25em]
\tableline\tableline\\
Total            & $1.6\pm0.6$ & $1.9$ (fixed) & \nodata & \nodata & $\phn3.4_{-1.9}^{+2.6}$                & 376.4(516) \\[2.5pt]
Optically Faint  & $1.8\pm0.6$ & $1.9$ (fixed) & \nodata & \nodata & $\phn4.3_{-2.3}^{+3.2}$                & 360.1(516) \\[2.5pt]
Optically Bright & $1.1\pm0.8$ & $1.9$ (fixed) & \nodata & \nodata & $\phn0.0_{-0.0}^{+2.0}$                & 145.3(516) \\[2.5pt]
J1159$+$0112     & $5.7\pm1.9$ & $1.9$ (fixed) & \nodata & \nodata & $11\phd\phn_{-5\phd\phn}^{+9\phd\phn}$ & 335.9(516)    
\enddata
\tablenotetext{a}{The temperature of the absorber is fixed at 30,000~$K$ for this model, and the Fe abundance is fixed at the solar value.} 
\tablenotetext{b}{The power law cut-off is fixed at 100~keV for this model, all abundances are fixed at the solar value, and the inclination angle is fixed at $\cos{i}=0.3$.}
\tablecomments{In addition to intrinsic absorption in units of $10^{22}~\cmsq$ (\hbox{${N}_{\rm H, int,22}$}), we included Galactic \nh\ \citep{dl90} in all models, and used Cash's statistics in the spectral fits. }
\end{deluxetable*}

\section{Hardness Ratios}\label{sec:hr}
Since we could only fit spectra for one object individually, we rely on measuring the X-ray hardness ratio to extract the intrinsic column density for the remaining BALQSOs.
We define a fractional hardness ratio, $HR = (H-S)/(H+S)$, where $S$ and $H$ are photon count rates in the observed soft (0.5--2.0 keV) and hard (2.0--7.0 keV) bands, respectively, for each source. Uncertainties were found using the Bayesian estimation method in \citet{park06}. We list the observed hardness ratio and associated uncertainties in Table~\ref{tab:hr}. 

\begin{figure*}
\begin{center}
\includegraphics[width=0.425\textwidth]{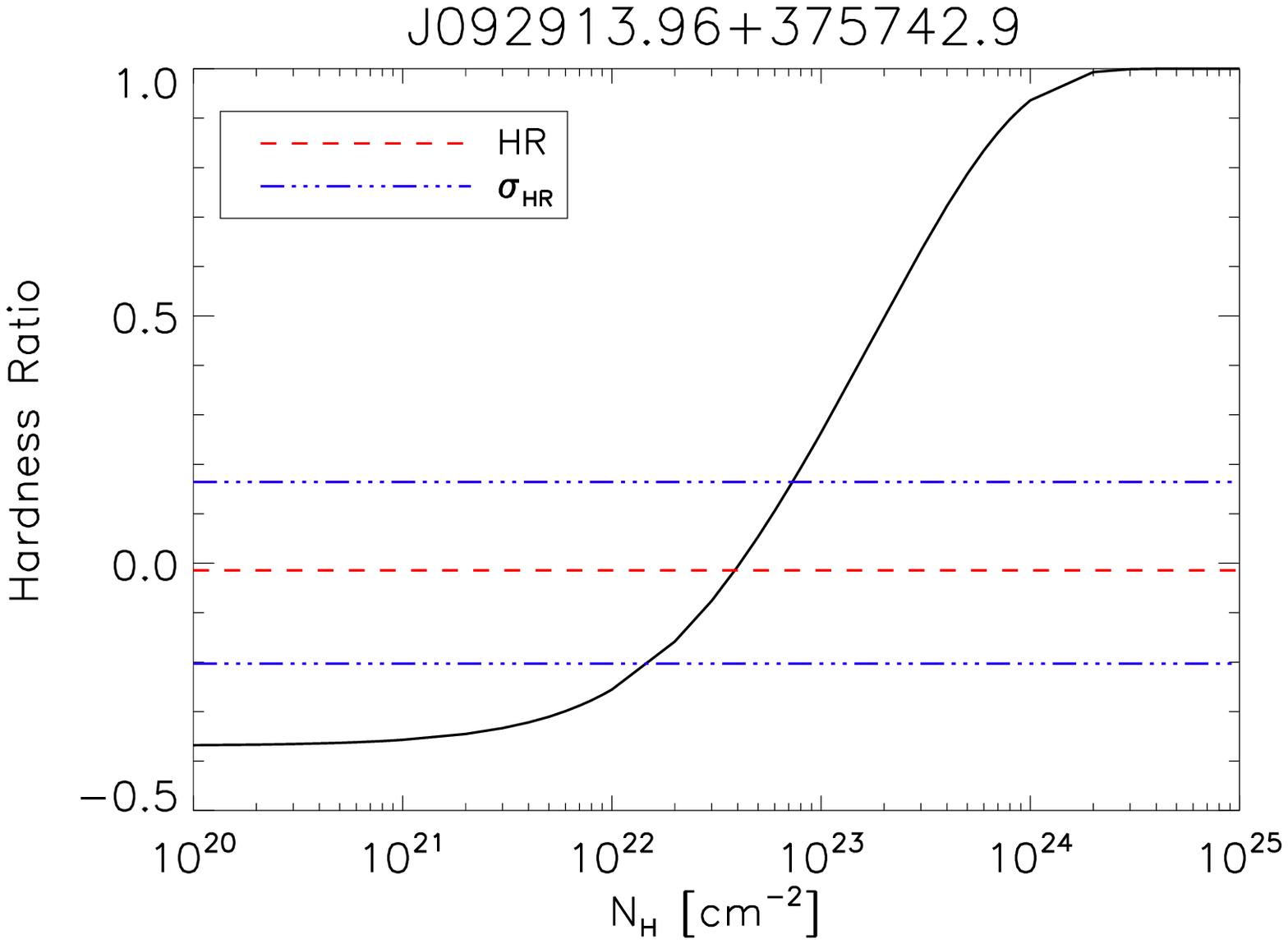}\\
\plottwo{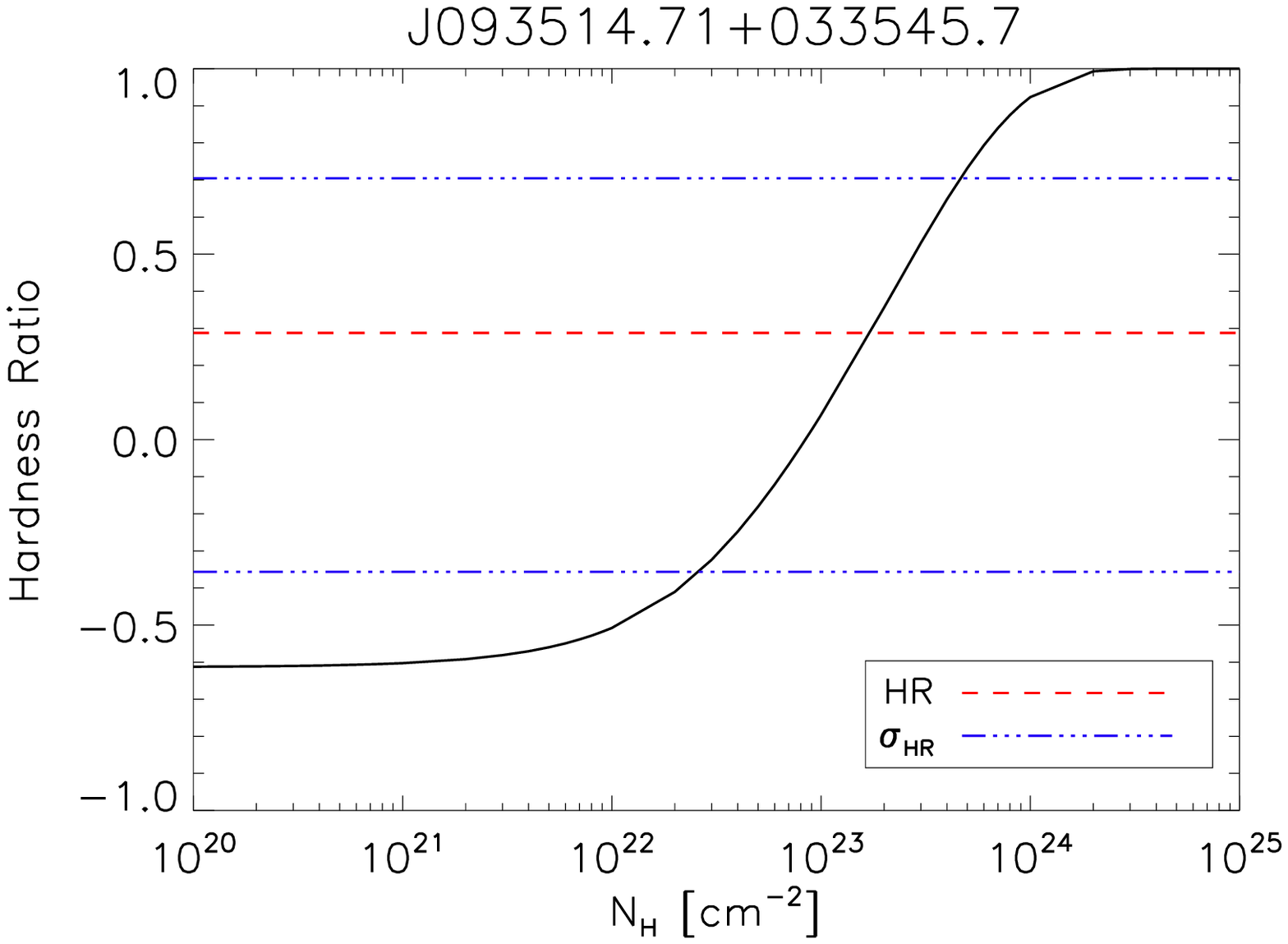}{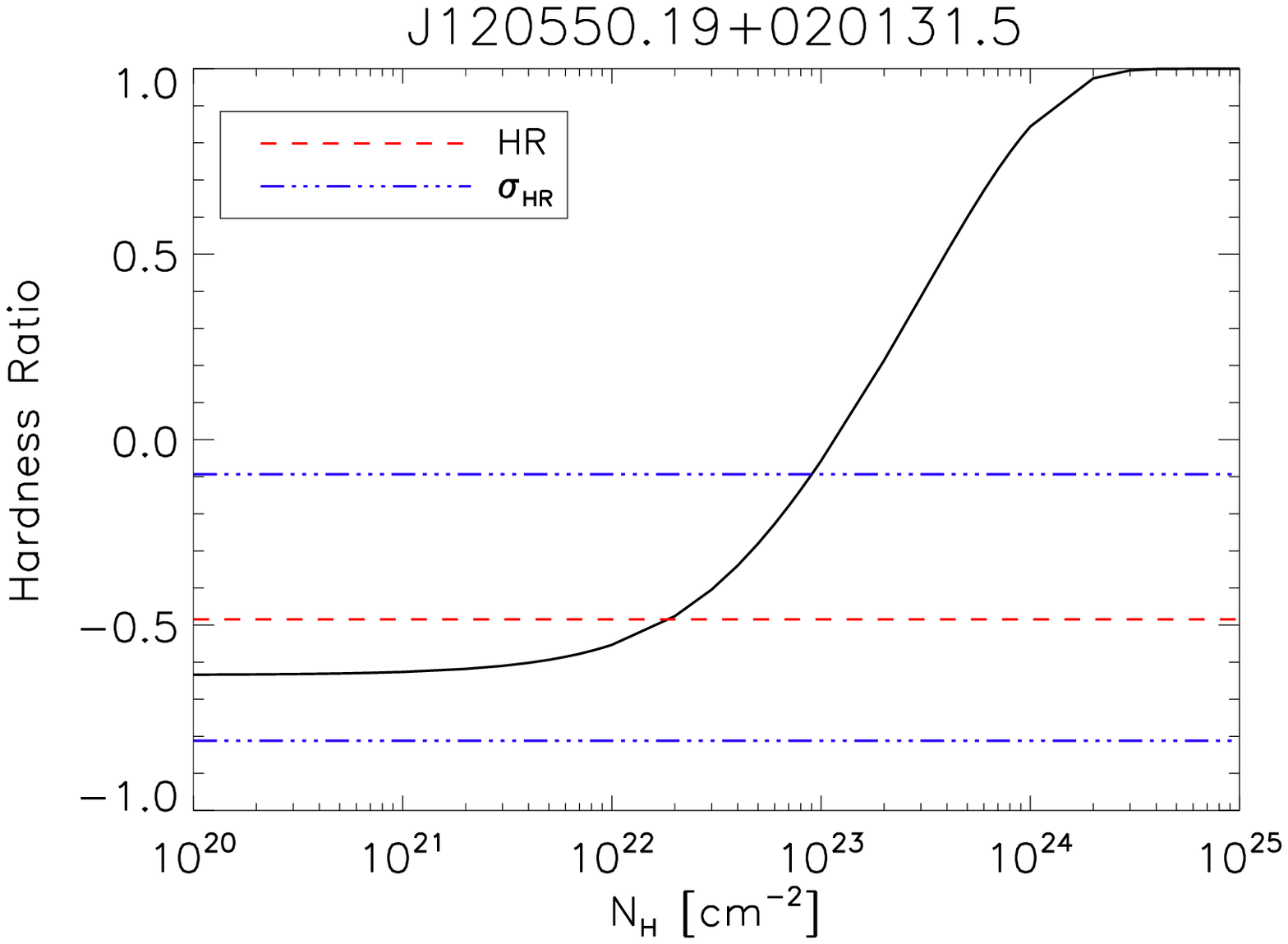}\\
\end{center}
\caption[Hardness Ratios]{Examples of predicted hardness ratios (solid black curves) plotted with the observed hardness ratio (dashed red line) and uncertainties (dot-dash blue lines). The measured HR and uncertainties provides a best fit $N_H$ and uncertainties. \label{fig:hr}}
\end{figure*}

Using the target-specific ancillary response and redistribution matrix files, model hardness ratios for a range of column densities were found with the aid of \verb+XSPEC+ models, using Model I (\verb+wabs*zwabs*zpowerlw+) as described in the previous section. For each observation, we explored the intrinsic column density in the range of 10$^{18-26}$ \cmsq . We used photon indices of $\Gamma = 1.54$ for the optically-faint sample, and $\Gamma = 2.10$ for the optically-bright sample, taken from the best-fit values in the spectral analysis. The un-normalized expected photon flux in the hard and soft bands was used to calculate the predicted $HR$ for each column density. We then compared the predicted to observed $HR$ to estimate the column density. Figure~\ref{fig:hr} shows three examples of how well this method works, and results for the complete sample are listed in Table~\ref{tab:hr}. The values for $N_H$ agree with what previous studies have found \citep[e.g.,][]{green01,gall99}.

In some cases, the uncertainties of the observed $HR$ are small enough that the column density is constrained within an order of magnitude. For other BALQSOs, the uncertainties are so large that the constraints span a couple orders of magnitude, while yet others have tight upper constraints but are unconstrained at the lower end, consistent with column densities of zero. The process breaks down for two objects: (1) J0959$+$6333, which does not have any photons in the hard band, providing HR $< -0.66$ (at the $1\sigma$ level); and (2) J1041$+$0001, which is dominated by the soft band, which contains over $\sim85\%$ of the photons, and has $HR=-0.40\pm0.15$. Neither of these objects have hardness ratios that are consistent with any model, and we do not include them in our further analysis.

\begin{deluxetable*}{lccccccl}
\tablewidth{0pt}
\tablecaption{Hardness Ratios and Column Density \label{tab:hr}}
\tabletypesize{\scriptsize}
\tablehead{\colhead{Object} &  \colhead{Photon flux} & \colhead{Photon flux} & \colhead{HR} & \colhead{$N_H$} & \colhead{$N_H$} & \colhead{$N_H$} & \colhead{Sample} \\
\colhead{} &   \colhead{(0.5--2.0 keV)} & \colhead{(2.0--7.0 keV)} & \colhead{Observed} & \colhead{Low} & \colhead{Best} & \colhead{High} & \colhead{} 
}
\startdata
J0046+0104 &   0.853$_{-0.453}^{+0.696}$  & 0.618$_{-0.394}^{+0.490}$ & $-0.160_{-0.504}^{+0.465}$ & 1$\times10^{18}$ & 2$\times10^{22}$ & 1$\times10^{23}$ & Opt. Faint \\[2.5pt]
J0929+3757 &   11.6$_{-2.95}^{+3.58}$ & 11.2$_{-2.43}^{+2.82}$ & $-0.0144_{-0.189}^{+0.179}$ & 1$\times10^{22}$ & 3$\times10^{22}$ & 7$\times10^{22}$ & Opt. Faint \\[2.5pt]
J0935+0335 &   2.29$_{-1.30}^{+3.00}$ & 41.3$_{-2.06}^{+2.94}$ & $0.288_{-0.644}^{+0.417}$ & 2$\times10^{22}$ & 1$\times10^{23}$ & 4$\times10^{23}$ & Opt. Bright \\[2.5pt]
J0959+6333\tablenotemark{a} & 10.7$_{-3.37}^{+5.32}$ & 0.00$_{-0.00}^{+1.83}$ & $-1.000_{-0.000}^{+0.341}$ & \nodata & \nodata & \nodata & Opt. Bright \\[2.5pt]
J1007+0532 &  14.9$_{-5.93}^{+8.42}$  & 13.8$_{-5.04}^{+6.51}$ & $-0.0356_{-0.336}^{+0.308}$ & 3$\times10^{22}$ & 1$\times10^{23}$ & 2$\times10^{23}$ & Opt. Bright \\[2.5pt]
J1010+5605 &  \nodata & \nodata & \nodata & \nodata & \nodata & \nodata & Opt. Faint \\[2.5pt]
J1041+0001\tablenotemark{b} &   7.92$_{-1.61}^{+1.85}$ & 3.35$_{-0.895}^{+0.985}$ & $-0.405_{-0.148}^{+0.149}$ & \nodata & \nodata & \nodata & Opt. Faint \\[2.5pt]
J1105+1115 &  0.542$_{-0.413}^{+0.832}$ & 0.344$_{-0.344}^{+0.557}$ & $-0.223_{-0.870}^{+0.849}$ & 1$\times10^{18}$ & 9$\times10^{21}$ & 2$\times10^{23}$ & Opt. Faint \\[2.5pt]
J1113+0914 &  \nodata & \nodata & \nodata & \nodata & \nodata & \nodata & Opt. Bright \\[2.5pt]
J1159+0112\tablenotemark{c} & 36.2$_{-5.58}^{+6.00}$ & 48.1$_{-5.52}^{+5.58}$ & 0.141$_{-0.099}^{+0.0945}$ & 3.8$\times10^{22}$ & 5.7$\times10^{22}$ & 7.6$\times10^{22}$ & Opt. Bright \\[2.5pt]
J1205+0201 &  6.49$_{-2.79}^{+4.03}$ & 2.25$_{-1.33}^{+2.09}$ & $-0.485_{-0.327}^{+0.391}$ & 1$\times10^{18}$ & 1$\times10^{22}$ & 9$\times10^{22}$ & Opt. Bright \\[2.5pt]
J1310+4822 &  2.69$_{-0.918}^{+1.18}$ & 2.07$_{-0.684}^{+0.786}$ & $-0.130_{-0.269}^{+0.251}$ & 1$\times10^{18}$ & 2$\times10^{22}$ & 5$\times10^{22}$ & Opt. Faint \\[2.5pt]
J1437+0428 &  \nodata & \nodata & \nodata & \nodata & \nodata & \nodata & Opt. Bright \\[2.5pt]
J1447+5203 &  5.73$_{-1.44}^{+1.72}$ & 5.02$_{-1.15}^{+1.26}$ & $-0.066_{-0.188}^{+0.176}$ & 1$\times10^{22}$ & 3$\times10^{22}$ & 6$\times10^{22}$ & Opt. Faint \\[2.5pt]
J1602+4013 &   0.864$_{-0.506}^{+0.814}$ & 0.653$_{-0.422}^{+0.543}$ & $-0.139_{-0.560}^{+0.498}$ & 1$\times10^{18}$ & 2$\times10^{22}$ & 1$\times10^{23}$ & Opt. Faint \\[2.5pt]
J1621+3555 &  1.76$_{-0.580}^{+0.822}$ & 1.24$_{-0.563}^{+0.660}$ & $-0.174_{-0.317}^{+0.305}$ & 1$\times10^{18}$ & 2$\times10^{22}$ & 7$\times10^{22}$ & Opt. Faint \\[2.5pt]
J2252$-$0841 &  1.44$_{-0.620}^{+0.896}$ & 0.766$_{-0.498}^{+0.609}$ & $-0.306_{-0.408}^{+0.410}$ & 1$\times10^{18}$ & 3$\times10^{21}$ & 6$\times10^{22}$ & Opt. Faint \\[2.5pt]
J2313+0034 &  \nodata & \nodata & \nodata & \nodata & \nodata & \nodata & Opt. Bright %\\[2.5pt]
\enddata
\tablecomments{~Photon flus is in units of 10$^{-6}$ phot cm$^{-2}$ s$^{-1}$.}
\tablenotetext{a}{J0959+6333 has a hardness ratio of $-1$ due to the lack of photons in the hard band.}
\tablenotetext{b}{J1041+0001 has an unconstrained column density due to low photon counts in the hard band.}
\tablenotetext{c}{The column density for J1159$+$0112 comes from spectral modelling assuming a reflected spectrum, not the hardness ratio of an absorbed power law.}
\end{deluxetable*}

\subsection{Intrinsic X-ray Weakness}\label{sec:xray}

To compare the intrinsic X-ray brightness to that of normal AGN,
we use the broadband spectral index \aox , which characterizes the UV-to-X-ray luminosity density ratio,
$\alpha_{\rm ox} = -0.3838\times\textrm{log}(l_{2500\text{\AA}}/l_{2 keV}).$
We follow the same method as in \citet{mora11}, calculating \aox\ for these objects and comparing the results with those unobscured quasars of similar luminosities. The rest-frame UV luminosity is calculated from 
the monochromatic flux at 2500\AA\ as reported by \citet{gib09a} for HiBALs (already corrected for Galactic extinction by the authors). To correct for the intrinsic extinction in the BALQSO host galaxy, we use the value $E(B-V)\approx0.023$ \citep{gib09a} which is thought to be typical for HiBALs. Since the dust reddening is expected to be SMC-like \citep{gib09a}, we use the value of $R_V = 2.93$ reported by \citet{pei92} for the SMC. Using the ranges of column densities found from the hardness ratios, we calculate the ranges of intrinsic X-ray luminosities and values of \aox\ for each BALQSO. 

\begin{figure*}
\plottwo{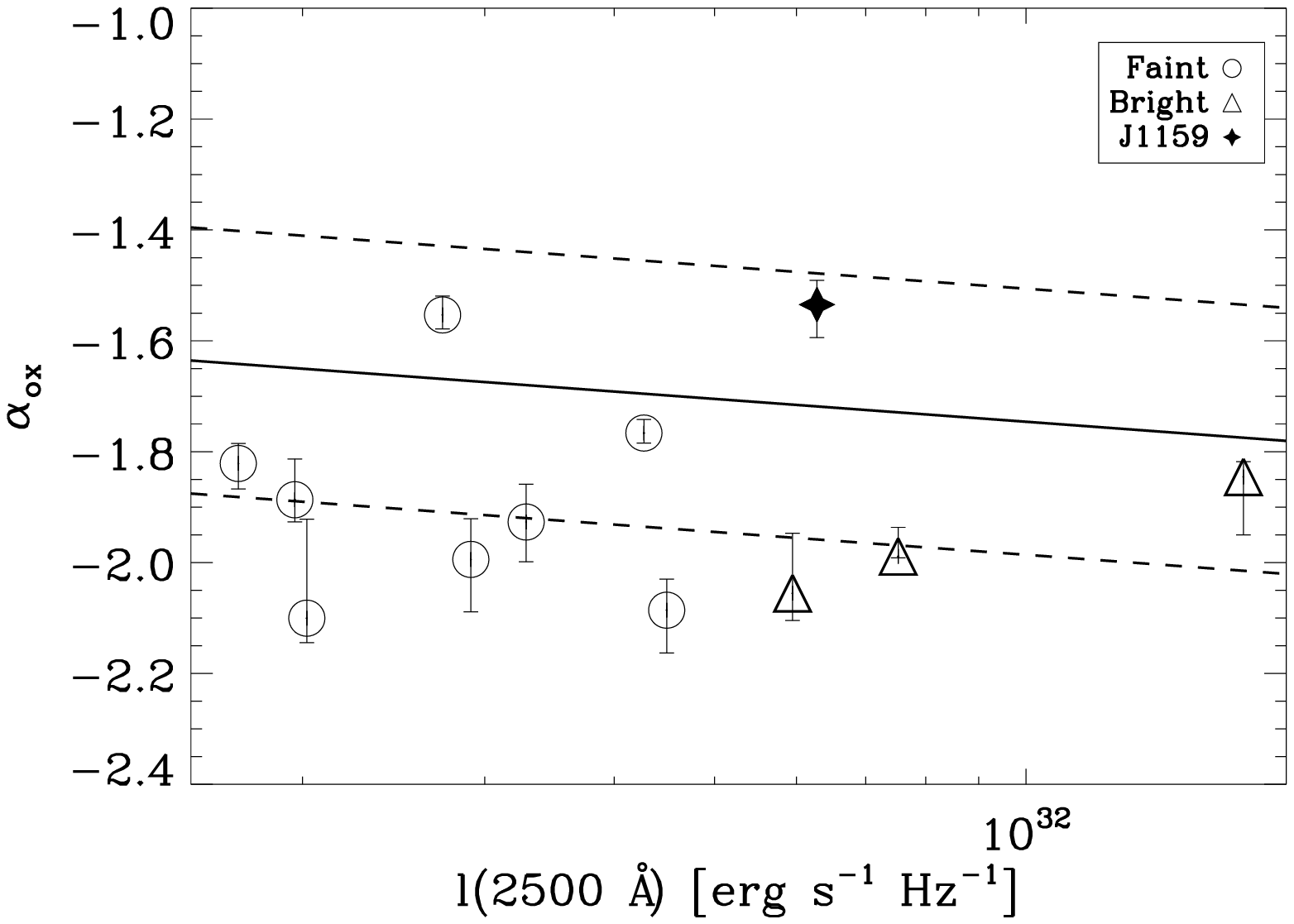}{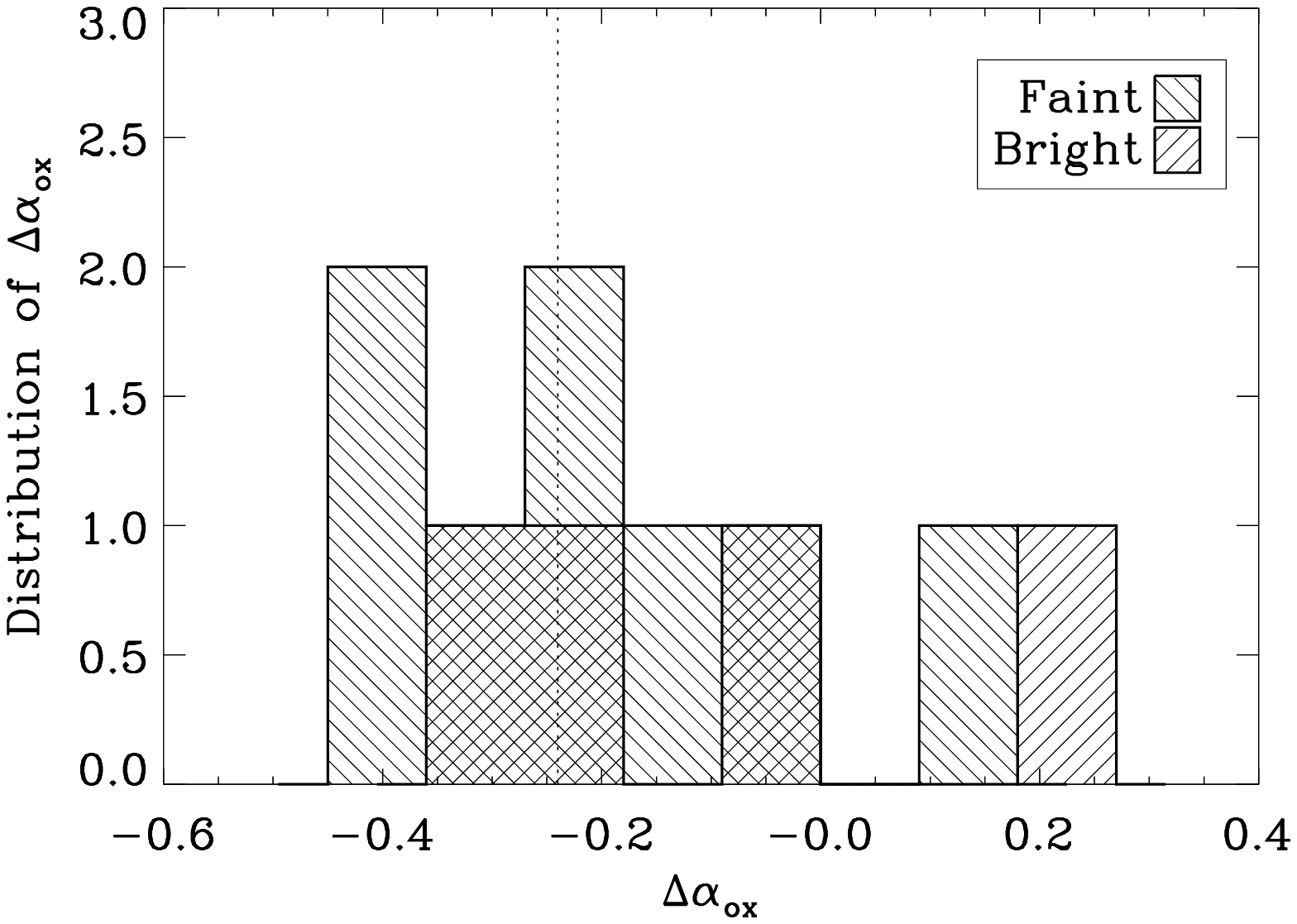}
\plottwo{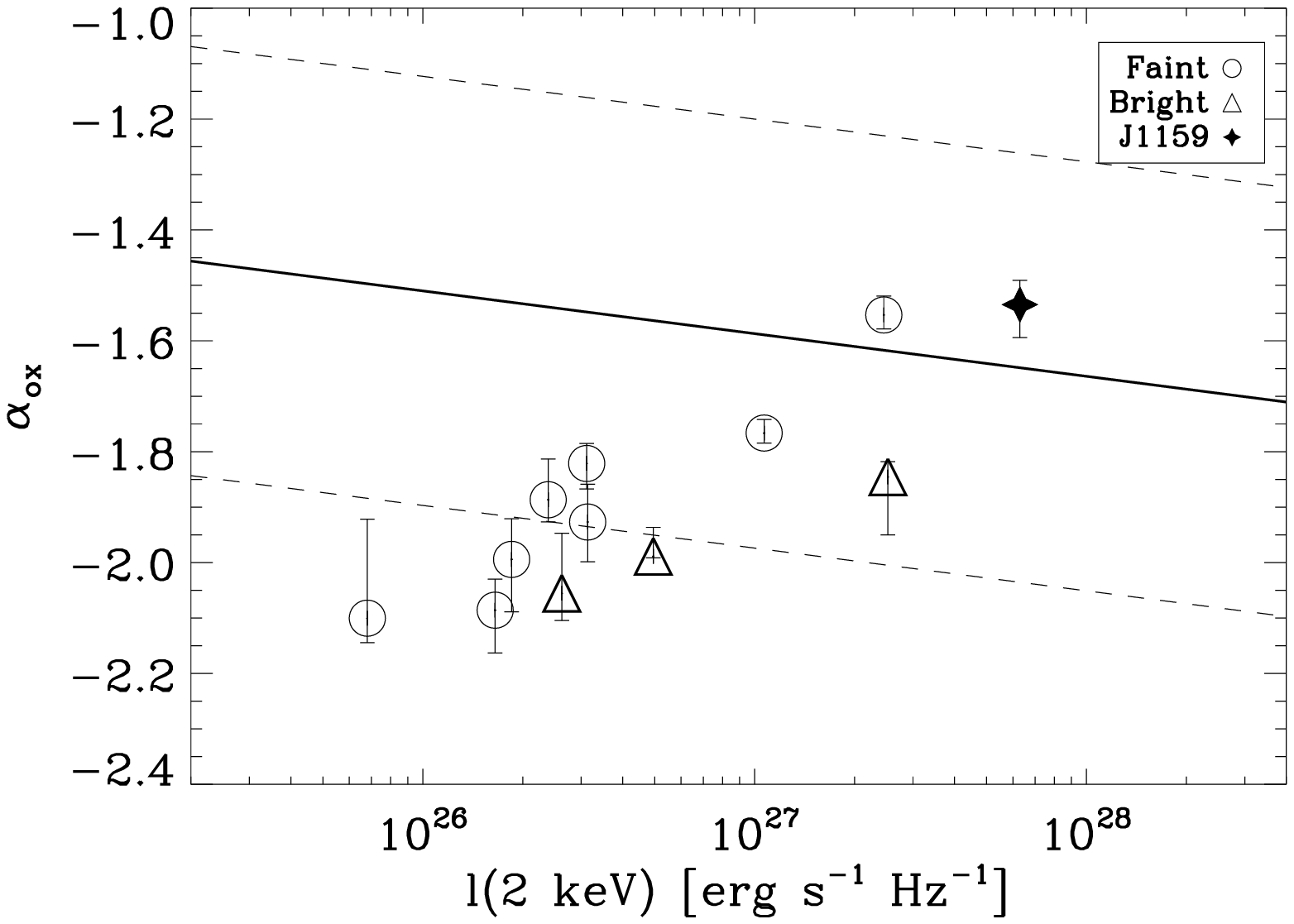}{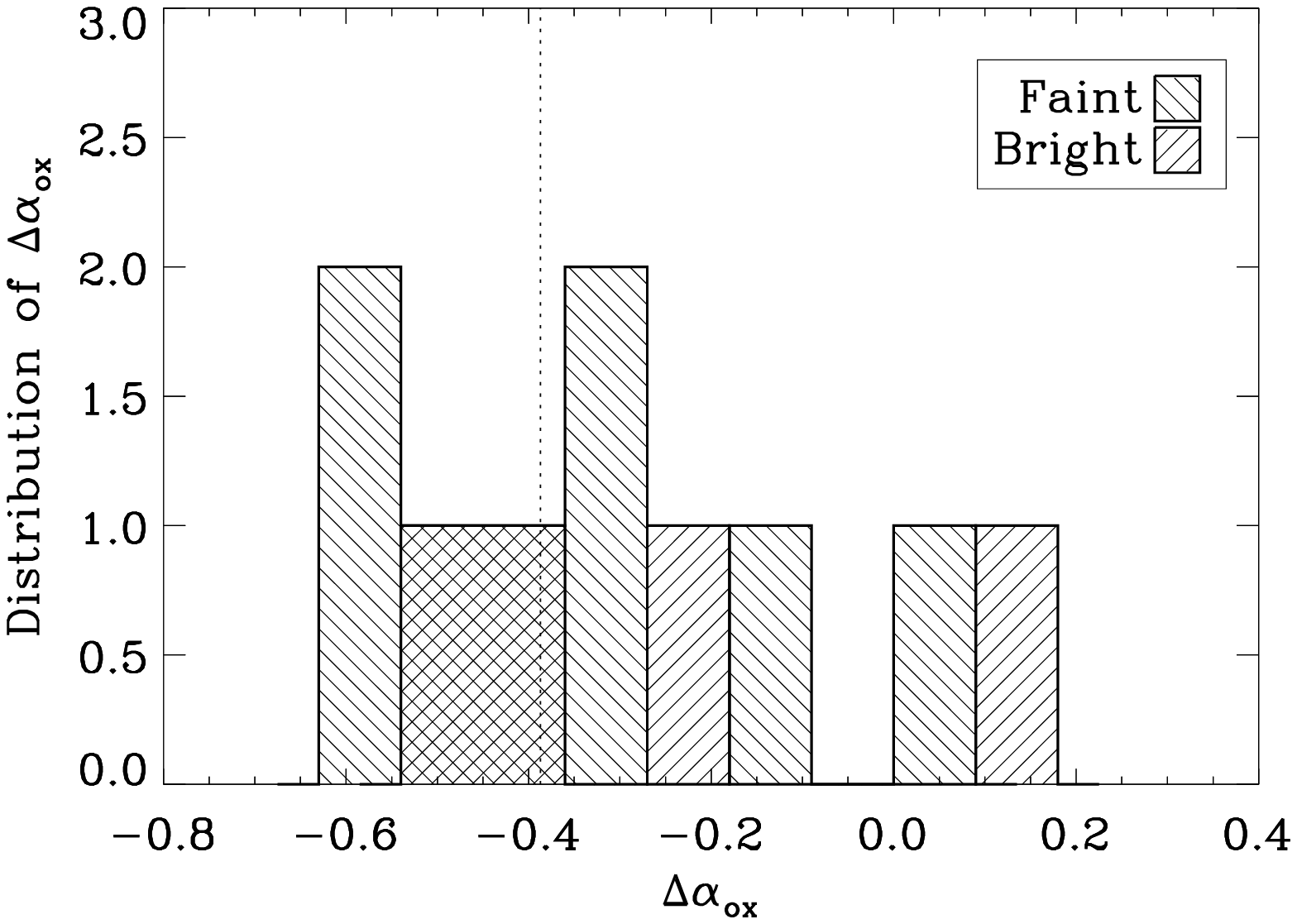}
\caption{Results for the UV--\aox\ relation are shown on the top, and the X-ray--\aox\ relation on the bottom, where \aox\ are intrinsic values after correcting for absorptions and extinctions. Luminosity for both relations versus \aox\ is shown on the left, with the relation for normal AGN plotted as a solid line and estimated S06 fit scatter are indicated by the dashed lines. The distributions of $\Delta$\aox\ are shown on the right, with the dotted lines indicating the S06 lower scatter boundary for the fit to a normal AGN sample. The symbols plotted are corrected for the best-fit $N_H$ from the hardness ratio analysis, and the error bars correspond to applying the upper and lower constraints placed on $N_H$ by the same. The circles and triangles distinguish the optically-faint and bright samples, respectively. In both cases, the samples are more intrinsically X-ray weak than expected. In particular, values of $\Delta$\aox\ are more negative for the optically-faint sample than the optically-bright sample, and both are more negative than normal AGN ($\Delta$\aox$=0$). \label{fig:uv}}
\end{figure*}

There are anti-correlations, \aox -- UV luminosity and \aox -- X-ray luminosity, observed in normal AGN. For example, Steffen et al. (2006; hereafter S06) fit a sample of 333 AGNs with an iterative least squares based method \citep{demp77}, which also estimates the weighted standard binomial scatter of the fit.  We compare the BALQSOs here with the S06 sample of normal AGN. In general we find fairly negative values of \aox\ for the optically-faint sample (see Figure~\ref{fig:uv}, top panels) compared to the \aox\ -- UV luminosity relation. Overall, 10 of the 12 objects are below the expected value for \aox . 
It is important to emphasize that these objects have already been corrected for intrinsic absorption and extinction, and therefore our results indicate that these objects are \textit{intrinsically} X-ray weak, rather than merely apparently X-ray weak. The \aox\ -- X-ray luminosity relation is weaker in AGN and has larger scatter (S06), but the preponderance of optically-faint BALQSOs are still well below the expected \aox , and only two objects are above, but still near, the expected value of \aox . 

The intrinsic X-ray weakness is better quantified by $\Delta$\aox , the difference between the measured \aox\ after correcting absorptions and extinctions and the expected \aox\ for unobscured AGN.
The histograms in Figure~\ref{fig:uv} show the values of $\Delta$\aox\ for optically-faint and optically-bright samples. 
The average $\Delta$\aox\ for the optically-faint and optically-bright samples is $-0.22\pm0.04$ and $-0.12\pm0.05$, respectively, for the \aox -- UV luminosity relation. While these fall within the S06 scatter ($\pm0.24$), the mean value of the optically-faint sample differs significantly from that of normal AGN ($\Delta$\aox $=0$). Using the \aox -- X-ray luminosity relation, we find that $\Delta$\aox\ for the faint and bright samples is $-0.34\pm0.04$ and $-0.26\pm0.05$, respectively, which again falls within the expected S06 scatter ($\pm0.387$).  
We fit linear relations between the intrinsic \aox\ values of BALQSOs and logarithms of UV or X-ray luminosity densities and obtained the following relations,
\be
\alpha_{\rm ox} = (-0.04\pm 0.21 )\times \log{\frac{l_{2500\AA}}{5\times10^{31}\lumind}} - 1.88 \pm 0.06
\\
\ee
\be
\alpha_{\rm ox} = (0.28 \pm 0.05) \times \log{\frac{l_{2keV}}{10^{27}\lumind}} - 1.79 \pm 0.03.
\ee

Most of these BALQSOs would require higher X-ray luminosities to make their intrinsic \aox\ values consistent with those of normal AGN. This could happen if \textit{(a)} the photon index is different such that we underestimated the intrinsic absorption or \textit{(b)} these objects are intrinsically X-ray weak compared to normal AGN. 

\begin{figure}
\plottwo{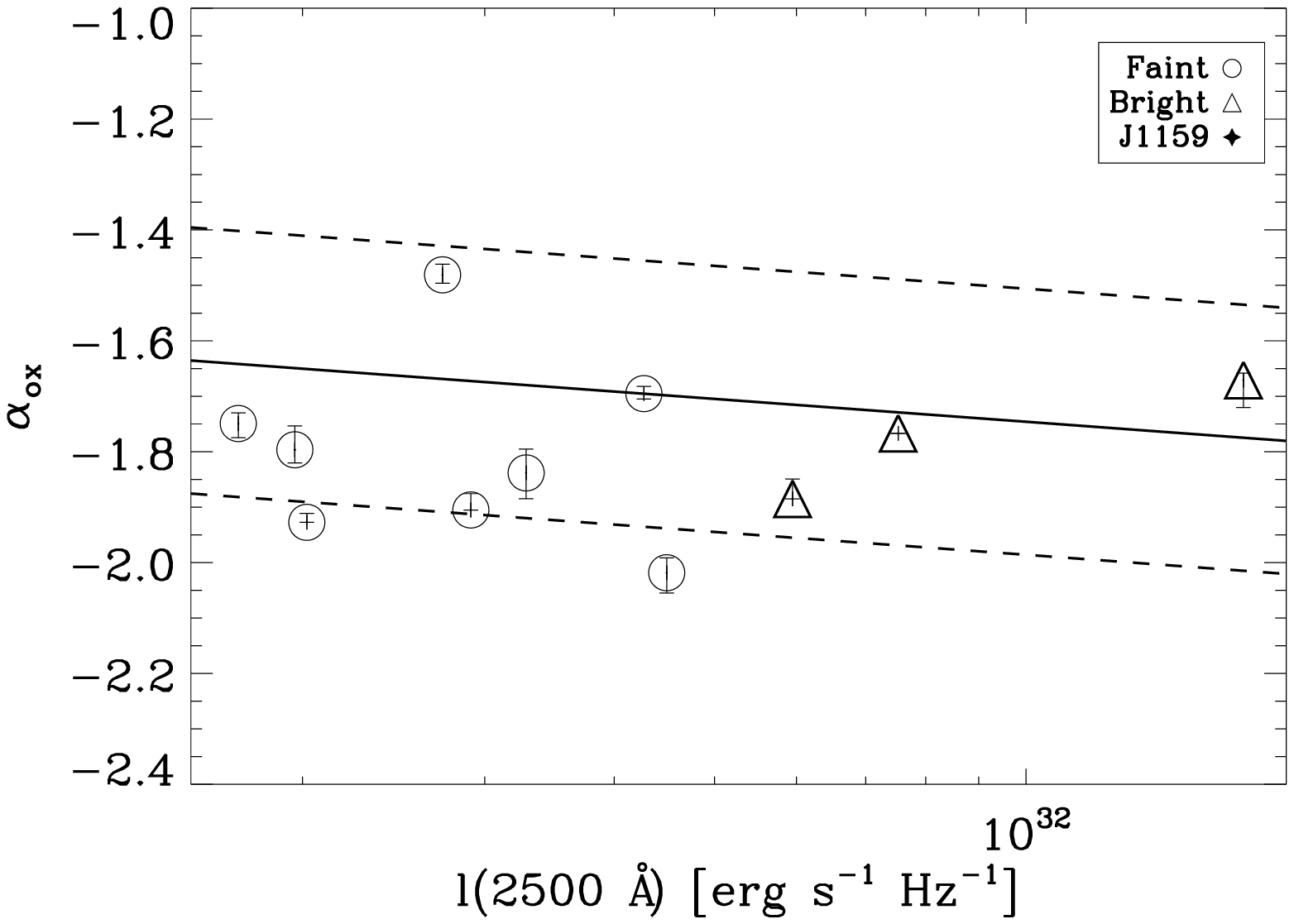}{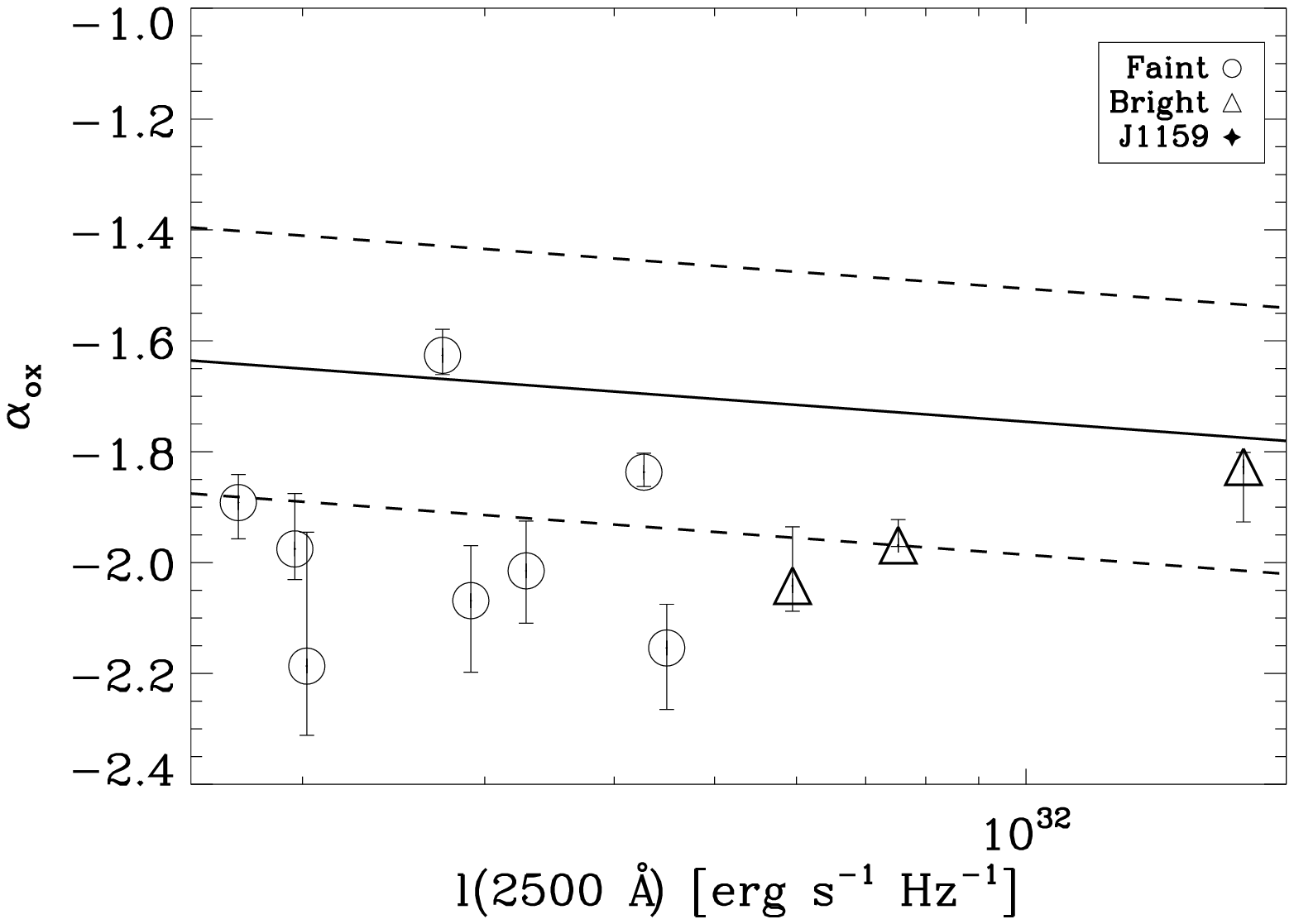}
\plottwo{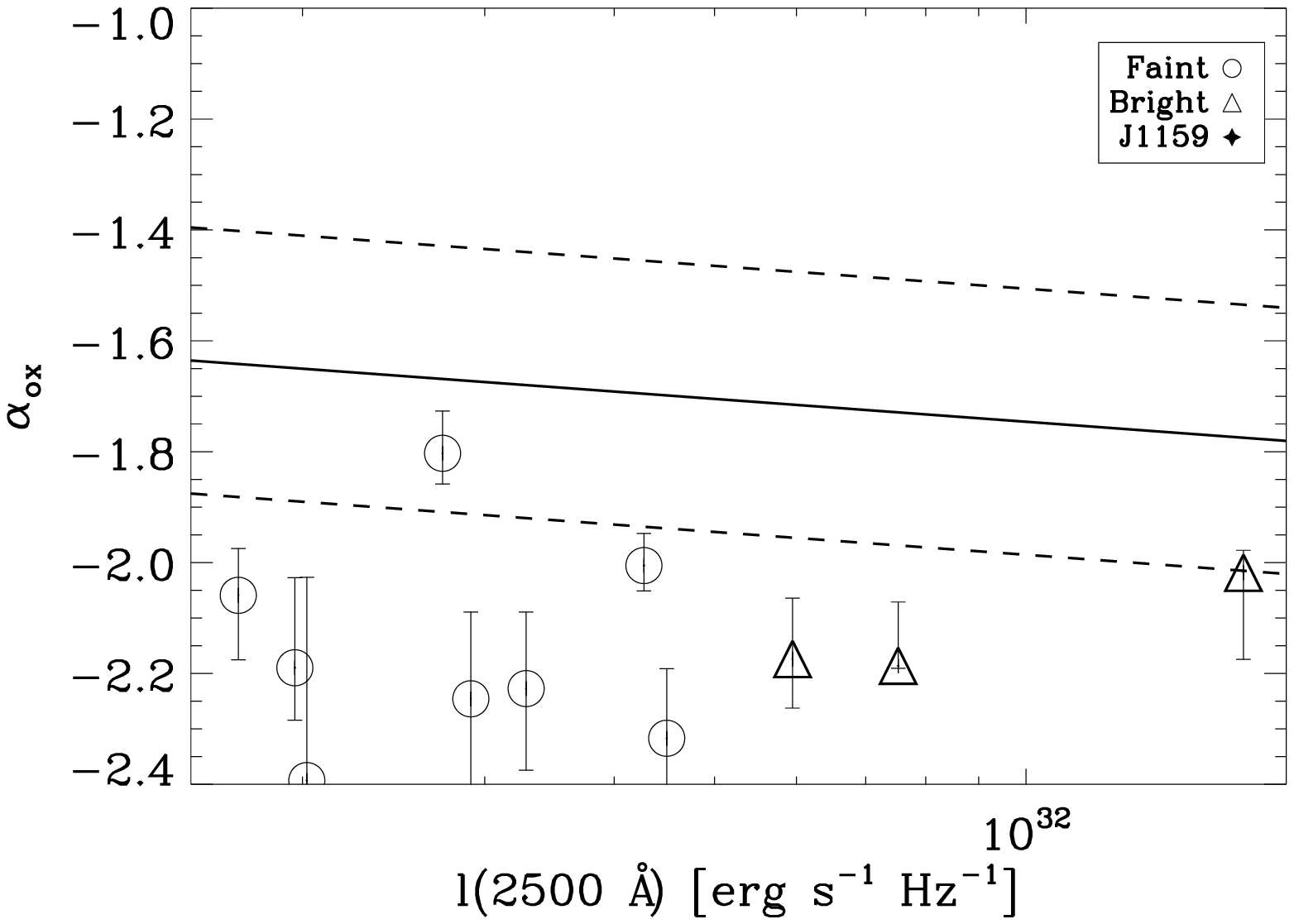}{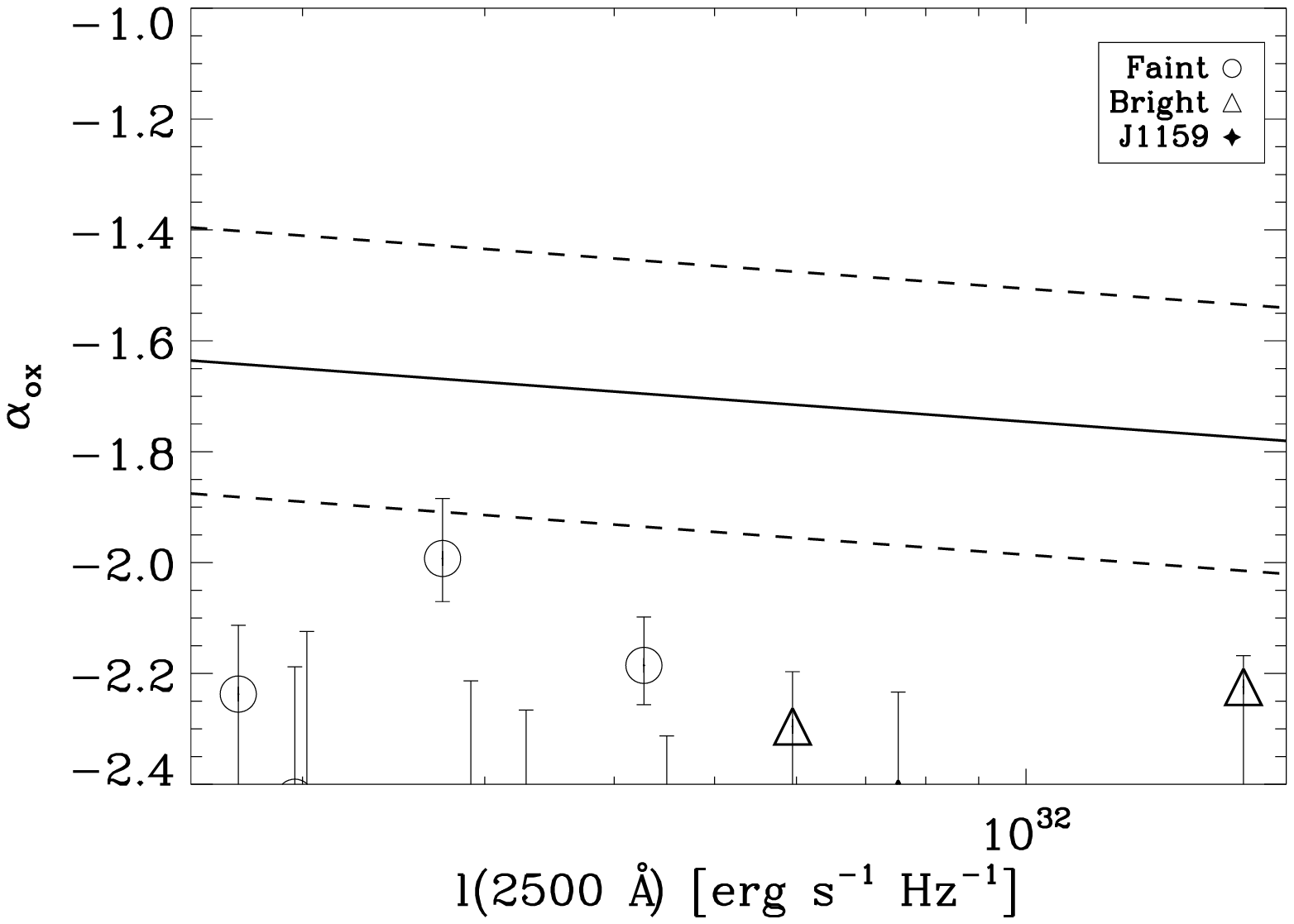}
\caption{We experimented with different photon indices, $\Gamma = 1$ (top-left), 2 (top-right), 3 (bottom-left), and 4 (bottom-right), used to constrain the intrinsic absorptions and absorption-corrected \aox\ values.  Only at $\Gamma=1$, an extremely flat photon index, the absorption-corrected \aox\ values fall in the expected range of normal AGNs.\label{fig:gamma}}
\end{figure}

To assess the first scenario, we run the same analysis but with different values of the photon index: $\Gamma=1,2,3,4$. We present the \aox -- UV luminosity relation seen in Figure~4a for each of these photon index values in Figure~\ref{fig:gamma}. 
The shallowest photon index, $\Gamma=1$, seems to provide values of \aox\ consistent with normal AGN, but this scenario has been shown to be unrealistic for quasars \citep{rt00,dai04,just07,saez08}. Steeper values of the photon index only serve to drive \aox\ to more negative values. Both samples here have the same photon index for each case, and we also tested adding values of $\Delta\Gamma=-1,+1,+2$ to the photon indices of $\Gamma=1.54$ and $\Gamma=2.10$ used for the different samples. The resulting trend is the same, with steeper photon indices providing more negative values of \aox , except that in no case are the samples in accordance with the expected scatter of normal AGN.

  We therefore rule out that the photon index is the cause, and conclude that these objects are intrinsically X-ray weak compared to normal AGN. Previous studies \citep[e.g.,][]{green01,gall06} have found these objects to be \textit{apparently} X-ray weak, but predict that the intrinsic SED should be that of a normal quasar, and therefore X-ray normal.

\subsection{Comparing the Optically-Bright and Faint Sample}\label{sec:civ}

We compare the absorption line properties between the X-ray detected optically-bright and optically-faint samples by  focusing on various properties of \ion{C}{4} absorption troughs to run statistical tests, taking values from \citet{gib09a} for the balnicity index (BI), modified BI (as defined in Gibson 2009), equivalent width, $v_{min}$, $v_{max}$, and $f_{deep}$ (the BAL bin fraction below 50\% of continuum) for the ion. Since the probabilities of the null hypothesis that the two samples were drawn from the same parent distribution was always above $36$\% ($P=0.36, 0.56, 0.63, 0.63, \text{ and }0.97$ for $v_{min}$, $v_{max}$, modified BI, $f_{deep}$, and BI, respectively), we find no strong evidence that the absorption line properties differ between the two samples. 

\begin{figure}
\begin{center}
\plotone{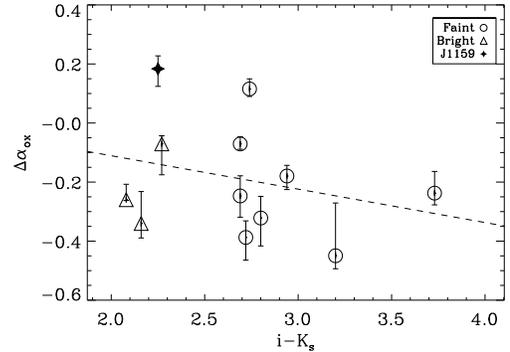}
\caption[$\Delta$\aox\ vs. $i-K_s$]{$\Delta$\aox\ plotted vs. $i-K_s$ colors for both the optically-faint and bright BALQSOs. The optically-faint BALQSOs dominate the right side of the plot and have a more negative mean value of $\Delta$\aox\ than those for the optically-bright BALQSOs. The dashed line is a linear fit to the data, and has a slope of $-0.11\pm0.13$. \label{fig:i-k}}
\end{center}
\end{figure}

We extended the comparison by including the constraints from the \chandra\ data. As seen from the \aox\ --UV relation, the optically-faint BALQSOs are more X-ray weak than the optically-bright ones, suggesting that the two samples may have different SEDs. We further check for any relation between $\Delta$\aox\ and the $i-K_s$ colors of the BALQSOs, as shown in Figure~\ref{fig:i-k}. We fit a linear relation and obtained
\be
\Delta\alpha_{\rm ox} = (-0.11\pm 0.13 )\times(i-K_s) + 0.12\pm 0.34,
\ee
providing a marginally negative slope, although the errors are still consistent with zero.

\begin{figure}
\begin{center}
\plotone{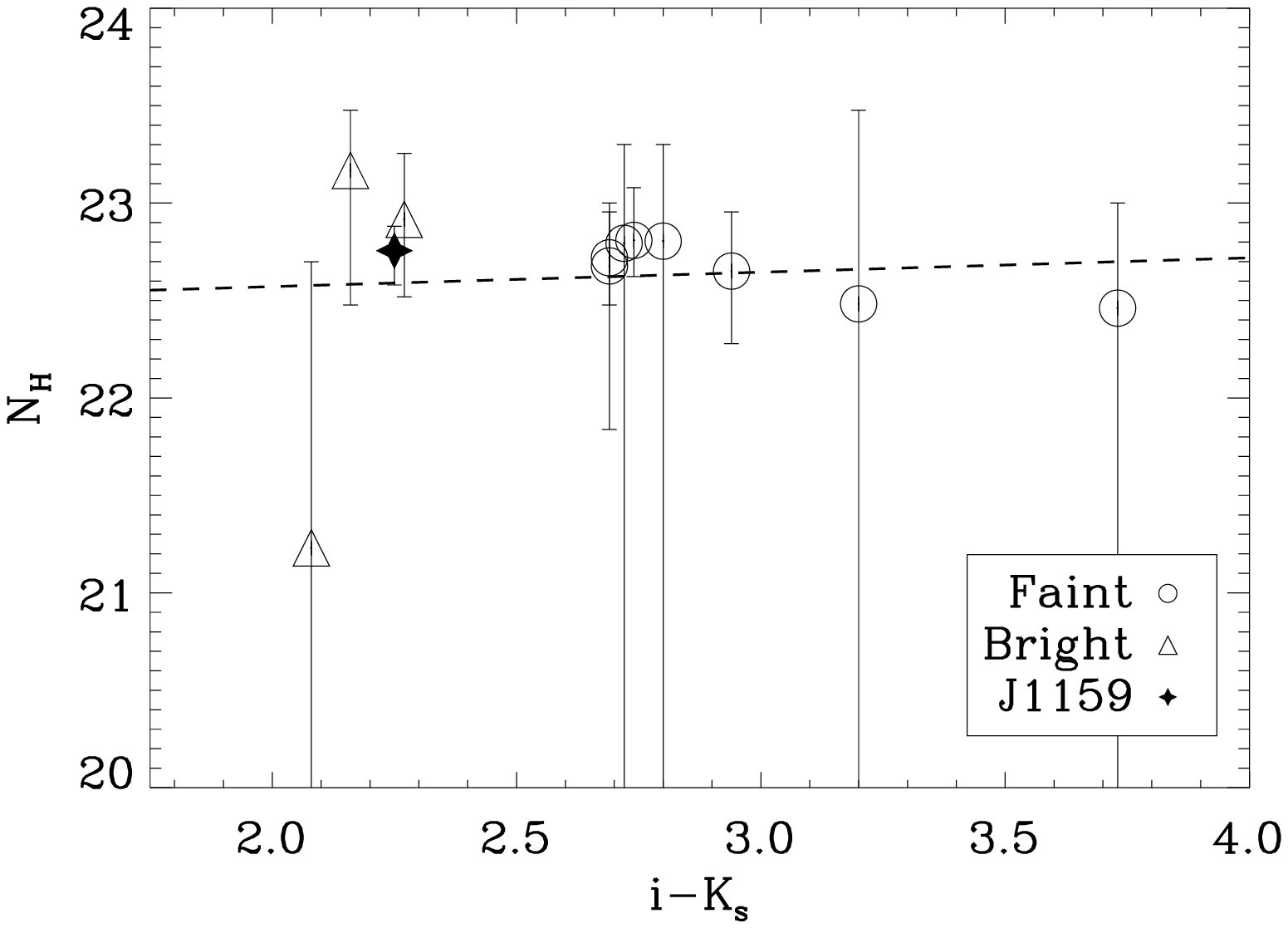}
\caption[$N_H$ vs. $i-K_s$]{$\textrm{log}N_H$ (with associated uncertainties) calculated from the hardness ratios is plotted against $i - K_s$. There is no clear difference between the two samples (the aberrant optically-bright BALQSO with a low best-fit value for $N_H$ has an upper error limit that is consistent with the rest of the sample). There is also no evidence of a relation between the two parameters, ﬁts to the data indicate $\log{\nh} \propto (0.074\pm0.32\times(i-K_s))$, as indicated by the dashed line. \label{fig:nh}}
\end{center}
\end{figure}

We finally consider column density, looking for a relation between $\log{N_H}$ and $i-K_s$ color (see Figure~\ref{fig:nh}). We use the value for $N_H$ calculated from the hardness ratios, along with the associated uncertainties. There does not appear to be a trend or a difference between the two samples. This is supported by a linear fit to the data, which yields 
\be
\log{N_H} = (0.07\pm 0.32)\times(i-K_s) + 22.42\pm 0.87. 
\ee
One optically-bright BALQSO (J1205+0201) has a value of $N_H$ that is $\sim2$ dex lower than the rest of the BALQSOs, but its upper limit is consistent with the rest of the sample. Therefore, any intrinsic difference between the two samples is unlikely to be due to column density.

\begin{deluxetable*}{lcccccccccc}
\tablewidth{0pt}
\tablecaption{Properties of 2MASS Selected BALQSOs\label{tab:par}}
\tabletypesize{\scriptsize}
\tablehead{\colhead{} & \colhead{} & \colhead{} & \colhead{} & \colhead{}      & \colhead{} & \colhead{} & \multicolumn{4}{c}{Kinetic Feedback} \\
           \colhead{}       &  \colhead{}          & \colhead{}            & \colhead{}             & \colhead{$M_{BH}$} & \colhead{}          & \colhead{}        & \multicolumn{2}{c}{$f_c$=20\%} & \multicolumn{2}{c}{$f_c$=43\%} \\
           \colhead{Object}       &  \colhead{$l_{2500}$}          & \colhead{$l_{2 keV}$}            & \colhead{$\Delta$\aox \tablenotemark{a}}             & \colhead{($10^8$\msun)}              & \colhead{$L_{bol}$}          & \colhead{$i-K_s$}        & \colhead{$v=0.1c$} & \colhead{$v=0.3c$} & \colhead{$v=0.1c$} & \colhead{$v=0.3c$}   \\
}
\startdata
J0046+0104                  & $2.9\times10^{31}$ & $1.9\times10^{26}$ & $-0.32_{-0.09}^{+0.07}$ & $9.93$ & $1.93\times10^{47}$ & $2.80$ &	0.057	& 1.54    &	0.12  &	3.3     \\
J0929+3757                  & $2.7\times10^{31}$ & $2.5\times10^{27}$ & $0.12_{-0.03}^{+0.03}$ & $9.95$ & $1.81\times10^{47}$ & $2.74$  &	0.065	& 1.76    &	0.14  &	3.79    \\
J0935+0335                  & $5.9\times10^{31}$ & $2.6\times10^{26}$ & $-0.34_{-0.05}^{+ 0.11}$ & $9.40$ & $3.95\times10^{47}$ & $2.16$ &	0.019	& 0.52    &	0.041 &	1.12    \\
J0959+6333                  & $8.3\times10^{31}$ & \nodata  & \nodata & $9.95$ & $5.49\times10^{47}$ & $2.06$ & \nodata & \nodata & \nodata & \nodata   \\
J1007+0532                  & $1.6\times10^{32}$ & $2.5\times10^{27}$ & $-0.07_{- 0.10}^{+0.03}$ & $10.32$ & $1.08\times10^{48}$ & $2.27$ &	0.032	& 0.87	  &	0.07  &	1.88    \\
J1010+5605                  & $2.3\times10^{31}$ & \nodata            & \nodata & $9.95$ & $1.56\times10^{47}$ & $3.44$ & \nodata & \nodata & \nodata & \nodata   \\
J1041+0001                  & $3.0\times10^{31}$ & $1.1\times10^{27}$ & \nodata & $9.95$ & $1.98\times10^{47}$ & $2.94$ & \nodata & \nodata & \nodata & \nodata   \\
J1105+1115                  & $2.0\times10^{31}$ & $6.8\times10^{25}$ & $-0.45_{-0.04}^{+ 0.18}$ & $9.95$ & $1.34\times10^{47}$ & $3.20$ &	0.041	& 1.12	  &	0.089 &	2.4     \\
J1113+0914                  & $6.8\times10^{31}$ & $4.0\times10^{25}$ & \nodata & $9.95$  & $4.54\times10^{47}$ & $1.84$ & \nodata & \nodata & \nodata & \nodata   \\
J1159+0112                  & $6.3\times10^{31}$ & $6.3\times10^{27}$ & $0.18_{-0.06}^{+0.04}$ & $9.95$ & $4.17\times10^{47}$ & $2.25$ &	0.025	& 0.67	  &	0.054 &	1.45    \\
J1205+0201                  & $7.5\times10^{31}$ & $4.9\times10^{26}$ & $-0.26_{-0.002}^{+0.05}$ & $10.75$ & $4.99\times10^{47}$ & $2.08$ &	0.0039	& 0.1	  &	0.0083&	0.22    \\
J1310+4822                  & $1.7\times10^{31}$ & $3.1\times10^{26}$ & $-0.18_{-0.05}^{+0.04}$ & $9.34$ & $1.15\times10^{47}$ & $2.94$ &	0.017	& 0.47	  &	0.037 &	1.01    \\
J1437+0428                  & $5.7\times10^{31}$ & \nodata            & \nodata & $9.95$  & $3.79\times10^{47}$ & $2.30$ & \nodata & \nodata & \nodata & \nodata   \\
J1447+5203                  & $4.3\times10^{31}$ & $1.1\times10^{27}$ & $-0.07_{-0.02}^{+0.02}$ & $10.16$ & $2.84\times10^{47}$ & $2.69$ &	0.055	& 1.48	  &	0.12  &	3.17    \\
J1602+4013                  & $4.5\times10^{31}$ & $1.7\times10^{26}$ & $-0.39_{-0.08}^{+0.06}$ & $9.95$ & $2.98\times10^{47}$ & $2.72$ &	0.038	& 1.03	  &	0.082 &	2.21    \\
J1621+3555                  & $3.3\times10^{31}$ & $3.1\times10^{26}$ & $-0.25_{-0.07}^{+0.07}$ & $9.76$ & $2.18\times10^{47}$ & $2.69$  &	0.026	& 0.7	  &	0.055 &	1.5     \\
J2252$-$0841                & $2.0\times10^{31}$ & $2.4\times10^{26}$ & $-0.24_{-0.04}^{+0.07}$ & $9.95$  & $1.31\times10^{47}$ & $3.73$ &	0.04	 & 1.09	  &	0.087 &	2.34    \\
J2313+0034                  & $1.5\times10^{32}$ & \nodata            & \nodata & $9.95$ & $9.78\times10^{47}$  & $2.43$ & \nodata & \nodata & \nodata & \nodata   %\\
\enddata
\tablenotetext{a}{$\Delta$\aox\ is the offset between the absorption and extinction corrected \aox\ and the \aox -- UV luminosity relation.} 
\tablecomments{~Luminosity densities are in units of \lumind\ and bolometric luminosities are units of \lumin.}
\end{deluxetable*}

\section{Discussion}\label{sec:dis}

Using the measured hardness ratios, we are able to constrain the column density for the entire detected sample to $N_H=3.5_{-2.6}^{+8.8}\times10^{22}$ \cmsq; for the optically-faint sample, the average is $1.9_{-1.7}^{+7.0}\times10^{22}$ \cmsq; for the optically-bright sample the average is $6.7_{-4.5}^{+12.5}\times10^{22}$ \cmsq.
However, only three out of 14 detected objects have column density measurements constrained within $\sim$1 order of magnitude. 
The value and range of $N_H$ are dependent on the photon index, since it dictates the shape of the model hardness ratio curves, seen in Figure~\ref{fig:hr}. If the model photon index steepens, the range of predicted $N_H$ will tighten, as the slope of the $HR$ steepens, by up to two orders of magnitude on the low density end. 

The UV luminosity densities, $\sim10^{31-32}$ ergs s$^{-1}$ Hz$^{-1}$, are on the high end for the sample of AGN in \citet{steffen06}, and while they could contribute somewhat to the more negative \aox\ values, they are not high enough to account for \aox\ values outside of the scatter seen in \citet{steffen06}.

The potential for additional physically motivated, complex spectral models present systematic uncertainties to the measured intrinsic \nh\ column densities and X-ray luminosities.  
Both a simple powerlaw model modified by intrinsic neutral absorption, and the more complex spectral models described in Section~\ref{sec:spec} produce statistically consistent fits to the stacked spectra. 
When compared to the model of simple powerlaw with neutral absorption, \nh\ measured using partial covering, warm absorber, and reflections models are 5.0, 7.8, and 1.5 times larger, respectively, using the stacked total spectrum as a reference. The higher column densities lead to increased intrinsic luminosities, and therefore \aox\ (which will become more positive by 0.08, 0.07, and 0.03, respectively). 
We assumed an average intrinsic extinction from SMC type dust in our analysis, and in the most conservative case of assuming no extinction correction, the \aox\ values will be more positive by 0.02. Using larger extinction corrections will make the intrinsic \aox\ even more negative.

The magnification of the X-ray emission via gravitational lensing implies that the true X-ray luminosity of the BALQSO will actually be weaker than we observe. \citet{chen13a} calculated the $\Delta$\aox\ differences between BALQSOs and normal quasars to be 0.02--0.2 for different spins and geometries, where BALQSOs are modelled close to the equatorial plane and are thus X-ray stronger if the underlying continuum is isotropic.  This introduces an average correction factor of $-0.1$ to the intrinsic \aox\ of BALQSOs. While this is not included in our plots and analysis above, the overall effect would be to make all values of \aox\ more negative. The projection effect, $\cos{\theta}$, where $\theta$ is the inclination angle, is roughly cancelled between optical and X-ray emission.  However, if the optical depth of the X-ray emission is moderate, the projection effect will be smaller for the X-ray emission, which will result in even higher X-ray luminosity viewed at large inclination angles and more negative correction factors when compared to non-BALs. Had the gravitational lensing effect been accounted for in previous studies, they would have concluded that BALQSOs are X-ray weak. More high S/N X-ray spectra from unbiased sample of BALQSOs will be essential to further strengthen the conclusions. Recently, {\emph{NuSTAR}} observations, which are less susceptible to the effects of absorption, of two local BALQSOs also found the BALQSOs were X-ray weak \citep{luo13}.

Previous studies of optically selected BALQSOs show that their intrinsic \aox\ values after correcting for absorption and extinction are consistent with normal AGN \citep[e.g.,][]{gall06}. However, because studies of X-ray properties of BALQSOs are made difficult by low photon statistics, previously the intrinsic column densities were estimated from $\Delta$\aox , and therefore the intrinsic \aox\ is already biased to be consistent with normal AGN. In this paper, we have instead measured \nh\ column densities independently using hardness ratios, and then calculate $\Delta$\aox\ based on correcting for the measured column densities.  
By comparing the calculated values of \aox\ with those of a large sample of AGN, we find that these objects are \textit{intrinsically} X-ray weak  after correcting for absorption. Using NIR-selected objects allows us to present these results for a relatively unbiased sample, which will more accurately reflect the overall properties of BALQSOs. 

Combining all the effects, we conclude that the intrinsic SEDs for BALQSOs are different from those of normal AGN, with lower intrinsic X-ray luminosities, i.e. that BALQSOs are \emph{intrinsically} X-ray weak.  

\subsection{AGN Kinetic Feedback Efficiency}
We estimate the kinetic feedback efficiency, $\epsilon_k = 2\pi\mu m_p f_c r \nh v^3/L_{Bol}$, of the X-ray absorbers in these BALQSOs using the column densities obtained from the X-ray hardness ratio. These efficiencies are listed in Table~\ref{tab:par}. 
Due to the low signal to noise ratios in our X-ray spectra, we are unable to measure or quantify all relevant parameters from this particular dataset, so we use other measurements to fill in the necessary parameters. 

As in \citet{mora11}, we assume that the wind is located between the UV and X-ray emission regions \citep[e.g.,][]{rogerson11,mora11}, at $\sim 40 r_g$, using the micro-lensing constraints of \citet{dai10}. 
We use virial black hole masses from for this sample \citep[estimated from line widths;][]{shen08}. For the four objects without a black hole mass estimate, we use an average of the measurements for the sample ($9.95\times10^8$\msun ).  
The velocity of the BAL wind is normally measured in the UV wind (up to $\sim 0.1c$), but since the X-ray absorbing wind is at a smaller radius, its velocity can be higher. This velocity has been measured from the blue-shifted X-ray absorption lines detected in a few gravitationally lensed BALQSOs (Chartas et al.\ 2002, 2003, 2007), and can reach 0.3--0.8$c$. 
Since the X-ray absorption lines are mostly detected in mini-BALs, it is possible this is viewed through the edge of the wind, where the wind can be fully accelerated. Thus, we consider the X-ray absorption line as providing an upper limit, and assume the wind velocity is 0.1--0.3$c$, between the estimates from the two methods. The covering fraction, $f_c$, ranges from 20--43\%, depending on which classification \citep{wey91,trump06} is used.  Using complex absorber models, such as partial covering or warm absorbers, will increase the column densities by a factor of several.
The last parameter, the bolometric luminosity, is found by extrapolating the UV luminosity to 5100\AA , using the mean spectral energy distribution for quasars from \citet{richards06}, and the same authors' bolometric correction of 10.3. 

\begin{figure}
\begin{center}
\plotone{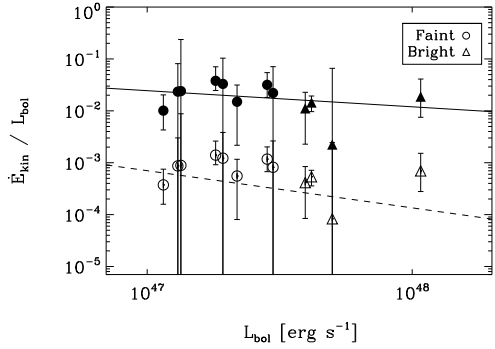}
\caption{The feedback efficiency vs.\ bolometric luminosity for BALQSOs, for a covering fraction of 43\%. The filled symbols are for a high wind velocity, $0.3c$, and the open symbols are for a low wind velocity, $0.1c$. Simple linear fits provide slopes consistent with zero for both samples. \label{fig:eff}}
\end{center}
\end{figure}

Although there are many assumptions involved in the calculation, these reflect our current understanding of the X-ray absorbing wind in BALQSOs. Other authors \citep[e.g.,][]{bo13} have been able to more accurately measure or quantify the feedback in the UV absorbers. Even though we can only provide numbers here that are tentative, this at least can provide general guidance on the magnitude of feedback energy from the X-ray absorbers to the community, 
prompting more accurate measurements in the future. Figure~\ref{fig:eff} presents order of magnitude estimates for the kinetic feedback efficiency for quasar winds as a function bolometric luminosity.
We find that low wind velocities produce low efficiencies: all are less than 1\%, regardless of the covering fraction. For high wind velocities, even at low covering fractions, we start to reach efficiencies over one percent. High wind velocities coupled with high covering fractions increase the efficiencies to a few percent, into the regime of kinetic AGN feedback. The average for high wind velocity, high covering fraction for the total sample is 2\%, almost half of the 5\% efficiencies required as described by \citet{silk98}. 
The optically-faint sample has a slightly higher average efficiency, 2.5\%, than the optically-bright sample, 1.2\%. With these wind velocities and covering fractions, it is possible that a feedback model such as that  proposed by \citet{he10} would be feasible. In this scenario, a ``two-stage'' feedback process requires efficiencies as small as 0.5\%, relying on a weak wind from the central engine to energize a hot, diffuse interstellar medium. This energized gas then amplifies the effect as it hits instabilities in a cold cloud within the host galaxy. 

\begin{figure}
\begin{center}
\plotone{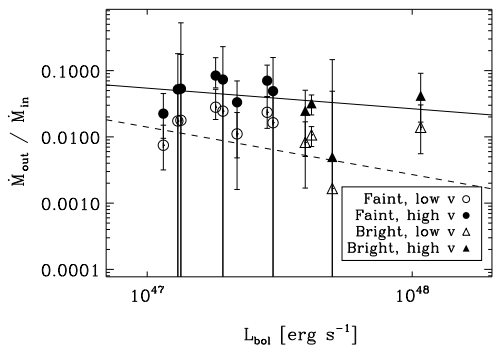}
\caption[$\dot{M}$ vs. $L_{bol}$]{The ratio of outflow to inflow $\dot{M}$ is plotted against the bolometric luminosity, for a covering fraction of 43\%. Open symbols show the values for low wind velocities, and filled symbols for high wind velocities.  Simple linear fits provide slopes consistent with zero for both samples.  \label{fig:inout}}
\end{center}
\end{figure}

We also calculate the ratio of mass inflow and outflow rate of these targets (Figure~\ref{fig:inout}).
The mass outflow rate is estimated by $\dot{M}_{out} = 4\pi\mu m_p f_c r \nh v$, and the inflow rate is estimated by $\dot{M}_{in} = L_{bol} / c^2 \eta$, assuming a typical value of $\eta=0.1$. 
The mass outflow rates for our targets are only a small fraction of the inflow rates, typically 1\%--10\%.
Thus, this mass outflow will not significantly alter the accretion process and the growth of the black hole.
This situation is in contrast to the outflow rates calculated in some Seyferts \citep[e.g.,][]{ck12}, where the outflow/inflow rate ratios were measured to be as high as $\sim 1000$, a challenging number to allow accretion.
We fit linear relations between kinetic feedback efficiencies or outflow/inflow ratios (assuming 10\% errors) as functions of $\log{L_{bol}}$ and obtain
\ba
\epsilon = (-0.01\pm 0.01) \times \log{\frac{L_{bol}}{3\times10^{47}\lumin}} \cr + 0.02\pm 0.003 
\\
\frac{\dot{M}_{out}}{\dot{M}_{in}} = (-0.03\pm 0.02) \times \log{\frac{L_{bol}}{3\times10^{47}\lumin}} \cr + 0.04\pm 0.01,
\ea
using high wind velocity, high covering fraction.

In general, we find that the X-ray absorbers in HiBALs can provide kinetic feedback energy consistent with the requirements from models of co-evolution between black holes and host galaxies. The less numerous LoBALs and FeLoBALs have higher \nh\ column densities \citep{mora11} than HiBALs. Since the intrinsic fraction of LoBALs and FeLoBALs are found to be much higher than previous estimates, $\sim 5$\% and $\sim 2$\% of the quasar population \citep{dai12}, it is interesting to observe that the product of covering fraction and column density, $f_c\nh$, is roughly a constant between HiBALs, LoBALs, and FeLoBALs.
Thus, if all other parameters are similar, such as the location and speed of the wind, LoBALs and FeLoBALs will provide similar feedback energy as HiBALs, and the total feedback energy would increase by a factor of three \citep[][ has estimated feedback efficiency for FeLoBALs to be up to $\sim7$\%]{mora11}.
In addition, UV winds in some quasars also provide significant feedback energy of up to 1--5\% \citep[e.g.,][]{moe09,bo13,arav13}. \citet{lucy13}, however, conclude that there is probably a selection bias in these objects that leads to larger distances and therefore larger kinetic luminosities.  It is generally thought that UV and X-ray absorbers are located at different places in the central engine \citep[e.g.,][]{ge07,mora11}. The upper limits of feedback efficiencies from UV and X-ray winds are remarkably similar; however, the UV winds are located much further out. If kinetic energy from both winds reach the host galaxies at large distances from the central engine, the contribution from each will be additive. However, it is possible that there are energy exchanges between the winds, and the contribution will be cumulative, as the higher speed X-ray winds boost the acceleration of the low speed UV winds (in addition to the usual radiation force from the UV continuum).

\section{Conclusion}\label{sec:conc}
We presented \chandra\ observations of 18 BALQSOs, ten of which were optically-faint, and eight of which were optically-bright. 
While the optically-bright BALQSOs are more likely to be optically selected, the optically-faint sample  are typically missing in previous studies.  
We used the X-ray hardness ratios to constrain the column density of these objects, with a mean value of $N_H\approx 3.5_{-4.2}^{+6.2}\times10^{22}$ \cmsq , in good agreement with previous studies. 
These column densities result in absorption corrected \aox\ values that are generally below the expected \aox\ from a large sample of normal AGN \citep{steffen06}, indicating that these objects are intrinsically X-ray weak. 
Considering the large flux magnification of X-ray emission by the central black hole at large inclination angles, the intrinsic X-ray emission of BALQSOs is even weaker.
Using complex absorption models, such as partial covering and warm absorbers, will increase the \aox\ of BALQSOs moderately, but cannot totally compensate for the X-ray weakness.  
We also find AGN kinetic feedback efficiencies on the order of several percent from the X-ray wind of HiBALs, indicating these objects can be good candidates for providing AGN kinetic feedback. 
The mass outflow rates are only small fractions of inflow rate such that normal accretion will not be significantly altered.
LoBALs and FeLoBALs, although rarer, can provide similar feedback energy per unit quasar.
The X-ray wind by itself appears to provide the required feedback energy in two-stage models of co-evolution between black holes and host galaxies, while the potential addition of feedback from the UV wind would suggest that the total BAL wind can provide the required feedback energy required in most models of co-evolution between black holes and host galaxies.

\acknowledgements
We acknowledge support for this work provided
by the National Aeronautics and Space Administration
through Chandra Award Number 
GO1-12130X issued by the Chandra X-Ray Observatory Center,
which is operated by the Smithsonian Astrophysical Observatory
for and on behalf of the National Aeronautics Space
Administration under contract NAS8-03060. FS acknowledges support from a Marie Curie grant. GRS acknowledges ﬁnancial support from an NSERC Discovery Grant.

\end{document}